\documentclass[11pt]{article}
\pdfoutput=1
\usepackage{amsmath, amsfonts, amssymb}
\usepackage{comment}
\usepackage{graphicx}
\usepackage{psfrag}
\usepackage{amsthm}
\usepackage[usenames,dvipsnames,svgnames,table]{xcolor}
\usepackage{enumerate}
\usepackage{tcolorbox}
\usepackage{mathtools}
\usepackage{soul}
\usepackage{subfigure}
\usepackage{mathrsfs}  
\usepackage{a4wide}
\usepackage{booktabs}
\usepackage{tikz}
\usepackage{tikz-cd}
\usepackage{makecell}
\usepackage{bbm}

\usepackage{color}
\definecolor{dark-gray}{gray}{0.20}
\definecolor{gray}{gray}{0.30}
\definecolor{light-gray}{gray}{0.80}
\definecolor{dark-red}{rgb}{0.7,0,0}
\definecolor{dark-green}{rgb}{0.1,0.4,0}
\definecolor{dark-blue}{rgb}{0.3,0.3,0.7}
\definecolor{light-blue}{rgb}{0.8,0.8,1}
\definecolor{swamp}{RGB}{240, 199, 197}

\usepackage{pifont}
\usepackage{setspace}

\newcommand{\be}{\begin{equation}}
\newcommand{\ee}{\end{equation}}

\def\be{\begin{equation}}
\def\ee{\end{equation}}
\def\bea{\begin{eqnarray}}
\def\eea{\end{eqnarray}}

\newcommand{\dd}{\mathrm{d}}

\newcommand{\iphi}{\text{i}_\Phi }
\newcommand{\tr}{\text{tr}}
\newcommand{\Str}{\text{STr}}

\def\simleq{\; \raise0.3ex\hbox{$<$\kern-0.75em
		\raise-1.1ex\hbox{$\sim$}}\; }
\def\simgeq{\; \raise0.3ex\hbox{$>$\kern-0.75em
		\raise-1.1ex\hbox{$\sim$}}\; }

\numberwithin{equation}{section}

\usepackage{jheppub} 
\hypersetup{
	colorlinks=true,
	linkcolor=dark-blue,
	citecolor=dark-red,
	urlcolor=dark-green,
	linktoc=page,
	pageanchor=false
}

\title{\centering Higher order corrections to KPV: \\ The nonabelian brane stack perspective\\
}

\author{Simon Schreyer$^1$}

\affiliation{$^1$ Laboratoire Charles Coulomb (L2C), Université de Montpellier, CNRS,\\ F-34095, Montpellier, France}
\emailAdd{simon.schreyer@umontpellier.fr}

\abstract{ In this work, we study the decay of $\overline{D3}$-branes in the setup of Kachru, Pearson, and Verlinde (KPV) at higher order in $\alpha'$ from the perspective of a nonabelian $\overline{D3}$-brane stack. We extend the leading order analysis of KPV by including higher order commutators as well as higher derivative corrections.
Recently, the KPV setup has been studied at higher order in $\alpha'$ from the NS5-brane perspective. It was found that in order to control $\alpha'$ corrections the quantity $g_sM^2$ determining the amount of warping in the Klebanov-Strassler throat has to be much larger than expected. This leads to serious issues when using the $\overline{D3}$-branes as an uplift to dS. The benefit of the analysis in this work is that the $\overline{D3}$-brane perspective is controlled when the distance between the branes inside the brane stack is substringy which is a regime not controlled on the NS5-brane side. As a main result, we find that the strong bound $g_sM^2\sim \mathcal{O}(100)$ obtained on the NS5-brane also holds in the regime accessible from the $\overline{D3}$-brane perspective. We also show that the novel way of uplifting proposed in the recent work on $\alpha'$ corrections to the KPV setup can only work for small warped throats. 
 
}

\begin{document}

\makeatletter
\let\old@fpheader\@fpheader

\makeatother

\maketitle


\section{Introduction}

A stack of $p$ $\overline{D3}$-branes at the tip of the Klebanov-Strassler (KS) throat \cite{Klebanov:2000hb} provides one of the best studied setups for metastable SUSY breaking within string theory when embedded into a compact Calabi-Yau with all moduli stabilized. As originally proposed in \cite{Kachru:2003aw}, the setup is therefore one of the leading candidates realizing a metastable dS vacuum in string theory.

It has been famously observed by Kachru, Pearson, and Verlinde (KPV) \cite{Kachru:2002gs} that $p$ $\overline{D3}$-branes at the tip of the KS throat can be classically unstable, and thereby unable to provide an uplift to dS. During this decay process, the $\overline{D3}$-branes puff up into an NS5-brane which subsequently annihilates with flux and forms a supersymmetric state which cannot be used as an uplift anymore. 

Due to the Myers effect \cite{Myers:1999ps}, there exist two dual descriptions of the KPV setup. On the one hand, the nonabelian brane stack perspective (which we will also refer to as the $\overline{D3}$-brane picture) explains the puffing up of the brane stack into a noncommutative, fuzzy two sphere with radius $R_{S^2}$.
On the other hand, one can study the KPV setup from the abelian perspective, the NS5-brane picture, in which the NS5-brane wraps the fuzzy $S^2$ and has $p$ units of worldvolume $F_2$ flux. Both pictures are useful since they are valid in different regimes of parameter space. The nonabelian picture is valid when the distance between the branes of the stack is substringy. In terms of $R_{S^2}$ this translates into $R_{S^2}\ll \sqrt{g_s\, p} \,l_s$ where $g_s$ the string coupling, and $l_s$ the string length.\footnote{The factor $\sqrt{g_s}$ is special to the KPV setup where one works in the S-dual frame.} The NS5-brane picture is valid when $R_{S^2} \gg l_s$. Depending on the value of $\sqrt{g_sp}$ both regimes can overlap. Therefore, studying both perspectives allows to access a broad regime of parameter space. 

If one wishes to use $\overline{D3}$-branes to uplift to dS it is a necessary condition to ensure classical stability of the $\overline{D3}$-branes. Classical stability of the setup is guaranteed at leading order in $\alpha'$ for $p/M<0.08$ \cite{Kachru:2002gs} where $M$ is the number of $F_3$ flux quanta on the $S^3$ at the tip of the throat. The bound can be translated into a bound on the quantity $g_sM^2$ by choosing
\begin{equation}
\label{eq:gm2leading}
    p=1\,, \qquad g_sM>1\,, \qquad p/M<0.08 \qquad \Longrightarrow \qquad g_sM^2>12\,,
\end{equation}
where $g_sM>1$ is imposed for supergravity control. 
The lower bound on the parameter $g_sM^2$ is phenomenologically very important. This can be understood by noting that the warp factor of the throat is given by $\sim \exp(-8\pi N/3g_sM^2)$ where $N$ is the contribution to the D3 tadpole from within the throat. For a working $\overline{D3}$-brane uplift one has to ensure that the uplifting term is as small as the pre-uplifted AdS minimum. This in turn requires large warping, i.e.~$N\gg g_sM^2$. But since the maximal D3 tadpole is limited by topology a lower bound on $g_sM^2$ is key. 
Applying this reasoning to KKLT \cite{Kachru:2003aw} or the LVS \cite{Balasubramanian:2005zx,Conlon:2005ki} leads to strong (or maybe even deadly) constraints. For references in the context of KKLT see \cite{Freivogel:2008wm,Carta:2019rhx,Gao:2020xqh,Bena:2020xrh} and in the LVS see \cite{Junghans:2022exo,Gao:2022fdi,Junghans:2022kxg,Hebecker:2022zme,Schreyer:2022len}.
Furthermore, $g_sM^2$ has to be chosen large enough to control the backreaction of the complex structure moduli such that conifold instabilities are avoided \cite{Bena:2018fqc,Blumenhagen:2019qcg,Bena:2019sxm,Randall:2019ent,Scalisi:2020jal,Lust:2022xoq}.\footnote{Note that another way how the conifold instabilities might be avoided is presented in \cite{Bento:2021nbb,ValeixoBento:2023nbv}. This comes at the cost of a large flux contribution to the D3 tadpole from the bulk region of the Calabi-Yau challenging tadpole cancellation.}

The phenomenological interest in the quantity $g_sM^2$ motivated the authors of \cite{Hebecker:2022zme,Schreyer:2022len} to study higher order $\alpha'$ corrections to the KPV decay process from the NS5-brane perspective and by that making the bounds \eqref{eq:gm2leading} on $p/M$ and $g_sM$ more precise. That $\alpha'$ corrections are crucial for determining the lower bound on $g_sM^2$ can be seen as follows. If $g_sM^2$ is small, the radius of the $S^3$ at the tip is typically small since $R_{S^3}\sim \sqrt{g_sM}$. But $\alpha'$ corrections (i.e.~curvature corrections) are suppressed by $R_{S^3}$ and hence will be large when $g_sM^2$ is small. On the contrary, if it would be possible to choose $g_sM^2$ arbitrarily large, $\alpha'$ corrections could be safely neglected and the tree level, metastable KPV vacuum would be consistent. But as explained above this is not possible due to the limited D3 tadpole. One should hence aim for the smallest possible value of $g_sM^2$ for which $\alpha'$ corrections are numerically suppressed. Since currently only $\alpha'^2$ corrections are known, the authors of \cite{Hebecker:2022zme,Schreyer:2022len} study their numerical size in the KPV setup to work out how large $g_sM^2$ actually has to be such that the $\alpha'$ corrected potential has a metastable minimum.\footnote{Note that this is a weaker condition than enforcing control over $\alpha'$ corrections as will be discussed below.}

The main result of \cite{Hebecker:2022zme,Schreyer:2022len} is then a bound on $g_sM^2$ taking all currently known $\alpha'^2$ corrections into account. The bound is given by $g_sM^2>144$ which is stronger by an order of magnitude compared to the leading order expectation \eqref{eq:gm2leading}. Crucially, the lower bound is realized in a regime where the radius of the $S^2$ which is wrapped by the NS5-brane is string size characterizing the very boundary of control of the NS5-brane analysis. Additionally, in the regime of string size $R_{S^2}$ the new way of uplifting proposed in \cite{Hebecker:2022zme} supposedly works.

This is precisely where this work ties in. Here, we study the KPV setup not from the NS5-brane perspective but from the perspective of a stack of nonabelian $\overline{D3}$-branes. The advantage of the nonabelian $\overline{D3}$-brane perspective is that it is controlled precisely when the distance between the branes inside the brane stack is small compared to the string length. This enables us to advance into the regime of parameter space where the radius of the NS5-brane in the minimum of the potential is string size. It is therefore possible to investigate the minimal bound on $g_sM^2$ and the new uplifting mechanism in this regime. 

We do so by extending the leading order calculation of the potential of a nonabelian $\overline{D3}$-brane stack of \cite{Kachru:2002gs} to include the next-to-leading order commutator corrections as well as the same set of $\alpha'$ corrections \cite{Bachas:1999um,Garousi:2009dj,Garousi:2010ki,Garousi:2010rn,Garousi:2011ut,Robbins:2014ara,Garousi:2014oya,Jalali:2015xca,Garousi:2015mdg,Jalali:2016xtv,BabaeiVelni:2016srs,Garousi:2022rcv} which has been encountered in \cite{Hebecker:2022zme,Schreyer:2022len} but here as corrections to $\overline{D3}$-branes.

Jumping ahead, let us quote the most important results.
In the region in $(g_sM,p/M)$ parameter space which is controlled by the $\overline{D3}$-brane stack calculations, we find the bound $g_sM^2 \approx \mathcal{O}(100)$. This matches the expectations from the NS5-brane perspective and extends the results to a different regime in parameter space. Further, we find that the uplifting potential stays strictly positive (and hence cannot be tuned arbitrarily close to zero by varying $g_sM$ and $p/M$) everywhere in $(g_sM,p/M)$ space except at the boundary of control where $\alpha'$ corrections become more important than the tree level contributions. This shows that the new way of uplifting (relying on the fact that the $\alpha'$ corrected uplifting potential can be tuned arbitrarily close to zero without requiring a large warp factor) proposed in \cite{Hebecker:2022zme,Schreyer:2022len} does not work in the large $g_sM$ and small $p/M$ regime. 

The rest of the paper is structured as follows. In Sect.~\ref{sec:kpv} we briefly review the KPV process including $\alpha'$ corrections from the NS5-brane perspective. In Sect.~\ref{sec:actions} we review the Myers action for a nonabelian brane stack and its most important features. This is essential for the calculation of higher order commutator corrections. In Sect.~\ref{sec:kpvd3} we derive and analyze the main results of this paper. We start by reviewing the KPV process at leading order from the perspective of a nonabelian brane stack in Sect.~\ref{sec:leadingorder}. Then in Sect.~\ref{sec:parametrics}, we explain the parametrics of higher order corrections from the perspective of a nonabelian brane stack. Finally, in Sects.~\ref{sec:alphacorr} - \ref{sec:pheno}, we first evaluate all known $\alpha'$ corrections to $\overline{D3}$-branes at the tip of the KS throat and then calculate higher order commutator corrections.
Further, we bring everything together and derive constraints for phenomenology. 
In Sect.~\ref{sec:comparison}, we compare the NS5-brane and the $\overline{D3}$-brane perspective and discuss limitations of our analysis. After concluding in Sect.~\ref{sec:conclusion}, we summarize various technical computations regarding $\alpha'$ corrections in App.~\ref{sec:app}.

\section{Summary of the KPV process from the NS5-brane perspective} \label{sec:kpv}

\begin{figure}[h]
    \centering
    \subfigure[]{\includegraphics[width=0.37\textwidth]{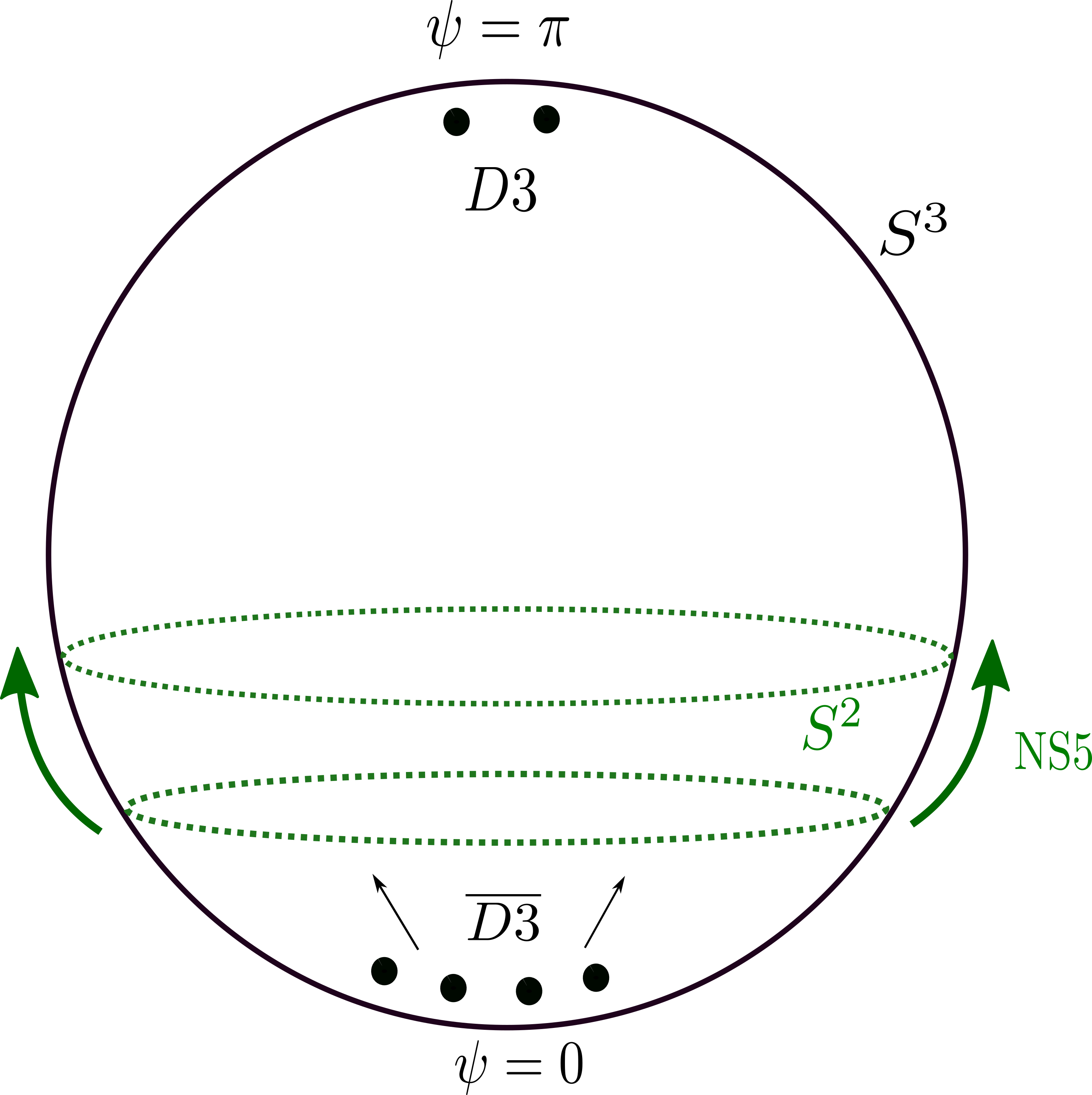}} 
    \subfigure[]{\includegraphics[width=0.62\textwidth]{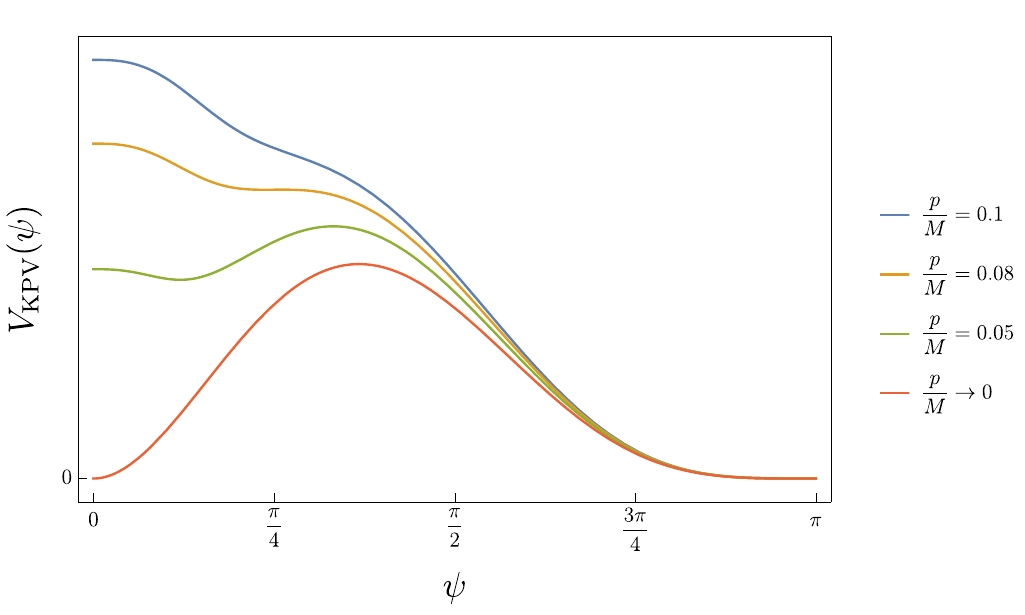}} 
    \caption{(a) illustration of the KPV decay on the $S^3$ at the tip of the KS throat. In (b) the leading order NS5-brane potential is shown for different values of $p/M$.}
    \label{fig:kpvreview}
\end{figure}

In the KPV setup \cite{Kachru:2002gs} a stack of $p$ $\overline{D3}$-branes is placed at the tip of the KS throat \cite{Klebanov:2000hb}. Geometrically, the KS throat is a deformed conifold with $M$ units of $F_3$ flux on the A-cycle of the conifold. At the tip the deformed conifold shrinks to an $S^3$ (which coincides with the A-cycle) with radius $R_{S^3}\sim \sqrt{g_sM \, \alpha'}$. 

The KPV decay process can now be understood in the following way. As we will show in detail below, for a stack of $\overline{D3}$-branes in an $F_3$ flux background it is energetically favorable to expand into a noncommutative fuzzy $S^2$ in the transverse directions \cite{Kachru:2002gs}. This state can also be described by an NS5-brane with $p$ units of worldvolume flux wrapping the $S^2$ inside the $S^3$ at the tip of the KS throat, see also Fig.~\ref{fig:kpvreview} (a) for an illustration of the KPV decay process. The potential of this NS5-brane along the $S^3$ is shown in Fig.~\ref{fig:kpvreview} (b), parametrized by the angle $\psi$ describing the position of the $S^2$ inside the $S^3$. From the potential one can read off that when $p/M\geq 0.08$, the NS5-brane slips over the equator of the $S^3$, annihilates with flux and forms a supersymmetric minimum described by $M-p$ D3-branes at the north pole ($\psi=\pi$) of the $S^3$ \cite{Kachru:2002gs}. This analysis is based on the leading order action of the NS5-brane.

As indicated in the Introduction the minimal value of $g_sM^2$ is a key phenomenological quantity. 
It is therefore important to test the leading order bounds on $p/M$ and $g_sM$ \eqref{eq:gm2leading} against $\alpha'$ corrections. This has been done in \cite{Hebecker:2022zme,Schreyer:2022len} by translating all currently known $\alpha'^2$ corrections to D-branes to the NS5-brane and subsequently evaluating them at the tip of the throat using the KS metric. The main result is the $\alpha'^2$ corrected potential of the NS5-brane enabling the study of the KPV decay at higher order.  
The potential is given by \cite{Schreyer:2022len}
\begin{equation}
\begin{split}
\label{eq:vtot}
    V_\text{tot} 
    = & \,\frac{4 \pi \mu_5 M}{g_s} 
      \sqrt{b_0^4 \sin^4 (\psi) + \left(p \frac{\pi}{M} -\psi +\frac{1}{2}\sin(2\psi) \right)^2}
       \times\Biggl[1+\frac{1}{(g_sM)^2} \Biggl( a_3 - a_1 \\
       & \qquad+ (a_4-2a_2) \cot^2\psi-a_2\cot^4\psi+\frac{ a_5 \cot^4\psi  }{ \sin^4\psi}\left(\frac{\pi p}{M}  -\left(\psi-\frac{\sin(2\psi)}{2}\right)\right)^2 \\
       & \qquad -\frac{a_6\,\cot^3\psi}{ \sin^2\psi}\left(\frac{\pi p}{M}  -\left(\psi-\frac{\sin(2\psi)}{2}\right)\right) \Biggr)\Biggr]\\
       & \qquad+\left[\frac{4\pi^2 p \mu_5}{g_s} -\frac{4 \pi \mu_5 M}{g_s} \left(\psi-\frac{\sin(2\psi)}{2}\right) \right] \left( 1+ \frac{a_7}{(g_sM)^2} + \frac{a_8 \cot\psi}{(g_sM)^2 \sin\psi}  \right)\,,
     \end{split}
\end{equation}
where $\mu_5$ is the brane tension, $g_s$ is the string coupling, $b_0 \sqrt{g_sM}\equiv R_{S^3}$ the radius of the $S^3$ at the tip, and $b_0^2\approx 0.93266$. The numerical constants $a_1$ to $a_8$ are given in \cite{Schreyer:2022len} but are not important in what follows. A detailed discussion of the potential can be found in \cite{Schreyer:2022len}.

\begin{figure}[h]
    \centering 
    \includegraphics[width=0.7\textwidth]{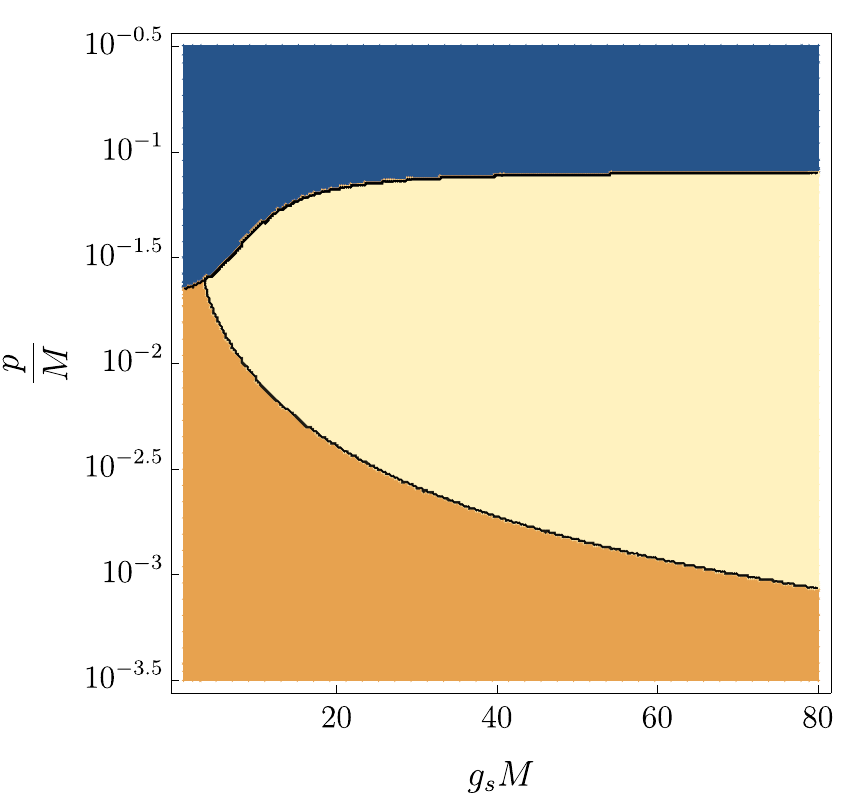} 
    \caption{Scan over the $(g_sM,p/M)$ parameter space taken from \cite{Schreyer:2022len}. In the blue region, no metastable minimum exist, in the yellow/orange region the minimum is at positive/negative energies.}
    \label{fig:kpv}
\end{figure}

The $\alpha'^2$ corrected bound on $p/M$ and $g_sM$ now follows by demanding that the potential \eqref{eq:vtot} should have a metastable minimum at positive energy in order to be able to use the setup for uplifting to dS. By scanning over the $(g_sM,p/M)$ parameter space one finds Fig.~\ref{fig:kpv} which depicts where in parameter space one has a minimum at positive (yellow) or negative (orange) energy. The $\alpha'^2$ corrected bound is then obtained by demanding $p/M$ and $g_sM$ in the yellow region. Importantly, the line where the metastable minimum is at zero energy is given by \cite{Schreyer:2022len}
\begin{equation}
    \frac{p}{M} \approx \frac{0.1029}{(g_sM)^{1.0909}}\,,
    \label{eq:zeroenergy}
\end{equation}
where $g_sM\geq 3.6$. The smallest value of $g_sM^2$ is obtained by choosing $g_sM$ as small as possible, i.e.~$g_sM=3.6$. With \eqref{eq:zeroenergy} and $p=1$ one finds 
\begin{equation}
    g_sM^2 \geq 144\,.
    \label{eq:boundalpha}
\end{equation}
One therefore concludes that $\alpha'$ corrections alter the leading order expectation $g_sM^2>12$ of KPV by an order of magnitude. 

The working assumption underlying \eqref{eq:boundalpha} is that the potential can be trusted in the yellow region of the $(g_sM,p/M)$ plane. As discussed in \cite{Schreyer:2022len}, this is only partially correct. The reason is that the main control parameter of the potential \eqref{eq:vtot}, the radius of the $S^2$ wrapped by the NS5-brane in the minimum of the potential, becomes string size close to the line of zero energy \eqref{eq:zeroenergy}. This signals the very boundary of control since higher order $\alpha'$ corrections become equally important. If one is looking for the minimal value of $g_sM^2$ where $\alpha'$ corrections can be neglected, the bound \eqref{eq:boundalpha} becomes even stronger.

The main goal of the following sections is to advance into precisely this region of parameter space where control is lost from the NS5-brane perspective. We do so by studying the KPV process from the perspective of a nonabelian stack of $\overline{D3}$-branes at higher orders in $\alpha'$ and commutators.

\section{Nonabelian D-brane actions }\label{sec:actions}

Before studying the KPV setup with a stack of $\overline{D3}$-branes at the tip of the KS throat, we briefly summarize the nonabelian D$p$-brane action \cite{Myers:1999ps} which is based on \cite{Douglas:1997ch,Douglas:1997zw,Douglas:1997sm,Hull:1997jc,Dorn:1996xk,Garousi:1998fg,Garousi:2000ea} and highlight its nonabelian features. For more details and explicit calculations we refer to \cite{Myers:1999ps,Myers:2003bw}.

As for a single D$p$-brane, the action of a nonabelian D$p$-brane stack has two parts: The DBI- and CS-action. They are given by\footnote{Greek indices run from $0\dots,D-1$ with $D$ the number of spacetime dimensions, the indices $a,b$ are D$p$-brane worldvolume indices and $i,j,k,\dots$ label directions normal to the D$p$-brane stack.} \cite{Myers:1999ps}
\begin{align}
\label{eq:dbi}
    \hspace*{-.3cm}S_\text{DBI} & = -T_p \int \dd^{p+1}\sigma \,\text{STr} \left(\text{e}^{-\phi}\sqrt{\det(Q^i_{~j})} \sqrt{-\det\left( P[E_{ab} + E_{ai}(Q^{-1}-\delta)^{ij} E_{jb}] +\lambda F_{ab} \right)}\right)\hspace*{-.3cm}\\
    \label{eq:cs}
    S_\text{CS} & \equiv T_p \int \text{STr} \left( P \left[ \text{e}^{i\lambda \iphi\iphi} \left( \sum C^{(n)} \wedge \text{e}^B \right)\right]\wedge\text{e}^{\lambda F}\right)\,,
\end{align}
with $T_p$ the brane tension, $\lambda=2\pi\alpha'$, $F_{ab}$ the worldvolume field strength, $B$ the Kalb-Ramond field, $C^{(n)}$ the RR $n$-form, $E_{\mu\nu}= G_{\mu\nu}+B_{\mu\nu}$, and $Q^i_{~j} \equiv \delta^i_{~j} + i\lambda [\Phi^i,\Phi^k]E_{kj}$. The matrix-valued scalar fields $\Phi^i$ describe the transverse displacement of the individual D-brane of the stack and can be identified with $x^i(\sigma)= 2\pi\alpha' \Phi^i(\sigma)$ where $\sigma^a$ are the worldvolume coordinates. The sum in \eqref{eq:cs} picks out the right $n$-form such that the integrand is a $(p+1)$-form. The main new property of the nonabelian action is that the scalars $\Phi^i$ are in the adjoint representation of the $U(N)$ gauge group of the brane stack (where $N$ is the number of branes) which makes the $\Phi^i$ in general noncommuting.
The nonabelian action reduces to the standard D-brane action when the $\Phi^i$ commute. 

This induces crucial new \textit{nonabelian features} in \eqref{eq:dbi} and \eqref{eq:cs} which do not exist for the abelian brane action. They are summarized by:
\begin{enumerate}
    \item All bulk fields are functions of the spacetime coordinates. Thus, they are implicitly functionals of the nonabelian scalars $\Phi^i$. The bulk fields therefore have to be \textit{nonabelian Taylor expanded} as e.g.
    \begin{equation}
        B_{\mu\nu} = \text{e}^{\lambda\Phi^i\partial_{x^i}} \left.B^0_{\mu\nu}(\sigma^a,x^i)\right|_{x^i=0} = \sum_{n=0}^{\infty}\frac{\lambda^n}{n!} \Phi^{i_1}\cdots \Phi^{i_n}\partial_{x^{i_1}}\cdots\partial_{x^{i_n}} \left.B^0_{\mu\nu}(\sigma^a,x^i)\right|_{x^i=0}\,.
    \end{equation}
    \item Each term on the brane has to be pulled back onto the worldvolume of the brane via the \textit{nonabelian pullback} $P$. It involves gauge covariant derivatives $D_a$ of the nonabelian scalars $\Phi^i$ and reads for instance for a two index tensor
    \begin{equation}
        P[E]_{ab} = E_{ab} + \lambda\, E_{ai} D_b \Phi^i +\lambda\, E_{ib} D_a \Phi^i +\lambda^2 E_{ij} D_a\Phi^i D_b\Phi^j\,.
    \end{equation}
    \item The object $\iphi$ in the CS-action denotes the \textit{nonabelian interior product} acting on a $p$-form $C^{(p)}=\frac{1}{p!} C^{(p)}_{\mu_1\cdots \mu_p} \dd x^{\mu_1}\wedge\cdots\wedge\dd x^{\mu_p}$ as
    \begin{equation}
        \iphi\iphi C^{(p)} = \frac{1}{2(p-2)!} [\Phi^i,\Phi^j] C^{(p)}_{ji\mu_3\cdots\mu_p} \dd x^{\mu_3}\wedge\cdots\wedge \dd x^{\mu_p}\,.
    \end{equation}
    \item The STr denotes the maximally symmetric trace \cite{Tseytlin:1997csa}. The trace is calculated by a symmetric average over all orderings of $F_{ab}$, $\Phi^i$, $D_a\Phi^i$ and $[\Phi^i,\Phi^j]$.\footnote{Note that before evaluating the symmetric trace, the bulk fields should be nonabelian Taylor expanded.} Hence, for the product of $n$ $\Phi$'s the STr reads
    \begin{equation}
        \text{STr}\left(\Phi^{i_1}\cdots\Phi^{i_n}\right) = \frac{1}{n!} \tr\left(\Phi^{i_1}\cdots \Phi^{i_n}+ \text{all perturbations} \right)
    \end{equation}
    This choice of trace matches the trace inferred from matrix theory \cite{Taylor:1999gq} and correctly reproduces string amplitudes to fourth order in $F_{ab}$ \cite{Tseytlin:1997csa,Tseytlin:1999dj} (which is enough for our purposes) but does not seem to capture the full physics of nonabelian fields at higher orders \cite{Hashimoto:1997gm,Myers:1999ps}.
\end{enumerate}

\section{The KPV process from the \boldmath{$\overline{D3}$}-brane perspective} \label{sec:kpvd3}

In this section, we derive and analyze the scalar potential for a stack of $\overline{D3}$-branes at the tip of the KS throat at higher order in $\alpha'$. These higher order corrections include commutator corrections derived from the nonabelian brane action and known $\alpha'$ corrections to D$p$-branes not included in the nonabelian action. Even though both types of corrections arise at higher order in $\alpha'$, we distinguish between them in the following and use the name $\alpha'$ corrections exclusively to refer only to all known higher derivative $\alpha'$ corrections to D$p$-branes. 

We start by briefly summarizing the leading order result of \cite{Kachru:2002gs} in Sect.~\ref{sec:leadingorder}. Then, in Sect.~\ref{sec:parametrics} we briefly discuss the parametrics of higher order corrections from the brane stack perspective. In Sect.~\ref{sec:alphacorr} we evaluate all currently known, non-vanishing $\alpha'$ corrections for a $\overline{D3}$-brane at the tip of the KS throat. Finally, in Sects.~\ref{sec:commutatorcorr} - \ref{sec:pheno} we derive the higher order commutator corrections and analyze the corrected scalar potential and its implications for phenomenology.

To set the stage, we introduce the DBI and CS action of a stack of $p$ $\overline{D3}$-branes in the S-dual frame as used by KPV \cite{Kachru:2002gs}. This in particular requires replacing $B_2$ by $-C_2$ and $B_6$ by $C_6$. The action is given by  
\begin{equation}
\label{eq:antid3}
\begin{split}
    \hspace*{-.3cm}S =& -\frac{T_3}{g_s} \int\dd^4\sigma \text{STr} \sqrt{\det\left( G_{ab} + 2\pi g_s F_{ab}\right)\det(Q) } 
      - T_3 \int  \text{STr}\Biggl(P\biggl[i\lambda\iphi\iphi (B_6 +  C_2\wedge C_4)\hspace*{-.3cm} \\
      &\qquad\qquad-\frac{\lambda^2}{2}(\iphi\iphi)^2 (B_6 \wedge C_2 + \frac{1}{2}C_2\wedge C_2\wedge C_4)+\dots\biggr]\Biggr)\,,
    \end{split}
\end{equation}
where
\begin{equation}
    \label{eq:Q}
    Q^i_{~j} = \delta^i_{~j} + \frac{i\lambda}{g_s}[\Phi^i,\Phi^k] \left( G_{kj} + g_s C_{kj}\right)\,.
\end{equation}
Some comments are in order.
In the DBI part of \eqref{eq:antid3}, we have neglected all terms under the square root which induce kinetic terms for the nonabelian scalars $\Phi^i$ (compare to \eqref{eq:dbi}). The kinetic terms will not matter for our purposes as we are only interested in stationary points of the potential where the kinetic terms vanish. 
As observed in \cite{Gautason:2016cyp}, it is possible to work in a gauge where $B_6$ is zero. By choosing a gauge where $C_4 = \dd\text{vol}_4/g_s$, one finds 
\begin{equation}
    H_7  = \frac{1}{g_s^2} \star_{10} H_3 = -\frac{1}{g_s} \, \dd\text{vol}_4 \wedge F_3 =  -  C_4\wedge F_3\,.
\end{equation}
Together with $H_7 = \dd B_6 + F_3 \wedge C_4$ this implies that $B_6$ can be set to zero. Note that it will turn out useful to work in this gauge to calculate the contributions of the CS action to the scalar potential. Further, we will set $F_2$ to zero since we are interested in the expansion of the $\overline{D3}$-brane stack in the directions transverse to the stack for which $F_2$ does not play a role.

\subsection{The leading order result} \label{sec:leadingorder}

Before including higher order corrections to the potential of a stack of $\overline{D3}$-branes let us summarize the leading order result of \cite{Kachru:2002gs}.  This is useful since when including higher order corrections we will go through precisely the same steps.

To derive the effective potential from \eqref{eq:antid3} one uses the nonabelian features listed in Sec.~\ref{sec:actions}.
The first step is to nonabelian Taylor expand the bulk fields $G_{\mu\nu}$, $C_2$, and $C_4$ which yields (abbreviating $C\equiv C_2$)
\begin{align}
    \label{eq:Gnonabelian}
    G_{\mu\nu}(\sigma^a,x^i) &= \eta_{\mu\nu}(\sigma^a,0)\,, \qquad \qquad  C_{4,\mu\nu\rho\sigma}=  C_{4,\mu\nu\rho\sigma}(\sigma^a,0)\,, \\
    \label{eq:Cnonabelian}
    C_{\mu\nu} &=  C_{\mu\nu}(\sigma^a,0) +\frac{\lambda}{3} \Phi^k F^{(3)}_{k\mu\nu}(\sigma^a,0)\,,
\end{align}
where we have made the assumption that the directions in which the brane stack expands into the fuzzy $S^2$ are flat. Ideally, to make closer contact with the KS background, one would like take into account the geometry of the tip of the throat. However, the flat approximation is valid when the radius $R_{S^3}$ of the $S^3$ at the tip is large compared to the string length.  
Using \eqref{eq:Gnonabelian} and \eqref{eq:Cnonabelian}, we can write \eqref{eq:Q} as 
\begin{equation}
\label{eq:QKPV}
    Q^i_{~j} = \delta^i_{~j} + \frac{i\lambda}{g_s}[\Phi^i,\Phi^k] \delta_{kj} + \frac{i\lambda^2}{3} [\Phi^i,\Phi^k] \Phi^l F^{(3)}_{lkj}\,.
\end{equation}
With this we can calculate the contribution of the DBI action to the effective potential up to order $\mathcal{O}(\lambda^2)$. One finds
\begin{equation}
    V_\text{DBI} = \frac{T_3}{g_s}\, \text{STr} \sqrt{\det Q^i_{~j}} = \frac{T_3}{g_s} \Biggl( p+\frac{\lambda^2}{4g_s^2}\tr\left( [\Phi^i,\Phi^j][\Phi^j,\Phi^i] \right) - \frac{i\lambda^2}{6} F_{ikl} \tr\left([\Phi^i,\Phi^k]\Phi^l\right)\Biggr)\,.
    \label{eq:vdbileading}
\end{equation}
The contribution from the CS action is given by evaluating the leading order term $\sim C_2\wedge C_4$ using the nonabelian features. This yields the same contribution as the cubic term in \eqref{eq:vdbileading} which is expected due to the imaginary self-dual flux background.\footnote{For a D3-brane the terms would cancel due to the no-force condition but for $\overline{D3}$-branes the terms add up.}
The total potential up to $\mathcal{O}(\lambda^2)$ is then of the form 
\begin{equation}
\label{eq:vonshellleading}
V_{\mathcal{O}(\lambda^2)} =\frac{T_3}{g_s} \Biggl( p+\frac{\lambda^2}{4g_s^2}\tr\left( [\Phi^i,\Phi^j][\Phi^j,\Phi^i] \right) - \frac{i\lambda^2}{3} F_{ikl} \tr\left([\Phi^i,\Phi^k]\Phi^l\right)\Biggr)\,.
\end{equation}
In order to see whether the brane stack expands into a fuzzy $S^2$ we study the minimum of the potential which is determined by the equation of motion
\begin{equation}
    0  = \frac{\lambda^2}{g_s^2} [[\Phi^n,\Phi^j],\Phi^j] - i\lambda^2 F_{njk} [\Phi^j,\Phi^k]\,.
    \label{eq:eomleading}
\end{equation}
As will be verified a posteriori, due to the $F_3$ flux background the $p$ $\overline{D3}$-branes will expand in their transverse directions into a fuzzy $S^2$ (which is the $S^2$ wrapped by the NS5-brane from the dual perspective) and it is therefore useful to make the following ansatz to solve \eqref{eq:eomleading}:
\begin{equation}
    [\Phi^i,\Phi^j] = A \,\varepsilon_{ijk} \Phi^k\,,
    \label{eq:ansatz}
\end{equation}
with $\Phi^k = -i A \alpha^k/2$. The $\Phi^k$ are proportional to the ($p \times p$)-dimensional, irreducible matrix representation $\alpha^k$ of the SU(2) algebra describing the fuzzy $S^2$. Its generators satisfy
\begin{equation}
    [\alpha^i,\alpha^j] = 2 i \varepsilon_{ijk} \alpha^k\,,
\end{equation}
with $i,j,k\in\{1,2,3\}$.
Assuming the $S^3$ to be large compared to the string length and since $F_3$ is constant on the $S^3$ at the tip we can approximate
\begin{equation}
\label{eq:f3nonabelian}
    F_{ijk } = f \varepsilon_{ijk}\,, \qquad\qquad \text{with} \qquad\qquad f = \frac{2}{b_0^3 \sqrt{g_s^3 M}}\,,
\end{equation}
where the expression for $f$ can be obtained from the quantization condition of $F_3$. The equation of motion is then solved for $A=-ig_s^2 f$ and the potential at the minimum is given by 
\begin{equation}
    V_{\mathcal{O}(\lambda^2)} = \frac{T_3 \,p}{g_s} \left( 1 - \frac{ \lambda^2g_s^6f^4(p^2-1)}{24}\right)\,,
\end{equation}
where we used that $\tr((\alpha^i)^2)=p(p^2-1)$. This energy is smaller than the energy of a stack of coincident $\overline{D3}$-branes\footnote{In this case all commutators vanish, i.e.~$A=0$.} such that the brane stack expands into a fuzzy $S^2$. 

Finally, using $\lambda=2\pi$ (we work in units where $\alpha'=1$), the radius of the fuzzy $S^2$ is given by
\begin{equation}
\label{eq:rfuzzy}
    R_{S^2}^2 \simeq \frac{\lambda^2}{p } \tr\left((\Phi^i)^2\right) = \frac{4\pi^2(p^2-1)}{b_0^8M^2} R_{S^3}^2\,,
\end{equation}
where $R_{S^3} = b_0 \sqrt{g_sM}$ is the radius of the $S^3$ at the tip. 

In our approximation, the background into which the brane stack expands is flat and hence does not reflect the actual geometry of the tip of the KS throat. It is therefore not possible to see the KPV decay explicitly but this is also not what we are aiming for from the brane stack perspective. We want to study the KPV setup when the fuzzy $S^2$ is string size while still keeping the $S^3$ much bigger to control the $\alpha'$ expansion. We will discuss this in more detail in Sect.~\ref{sec:comparison}.
One can nevertheless derive a useful constraint for when the potential can have a classical, SUSY breaking minimum, namely only if $R_{S^2} \ll R_{S^3}$. Using \eqref{eq:rfuzzy}, this translates into
\begin{equation}
    \frac{p}{M} \ll \frac{b_0^4}{2\pi} \approx 0.138\,.
    \label{eq:treelevelconstraint}
\end{equation}
This constraint agrees, albeit being a bit weaker, with the constraint $p/M<0.08$ from the NS5-brane perspective.

In the next sections we will redo this calculation taking into account higher order corrections into the potential of the nonabelian scalars $\Phi^i$. We do so by expanding the nonabelian action \eqref{eq:antid3} up to $\mathcal{O}(\lambda^4)$ and by including all known $\alpha'$ corrections to D$p$-branes into our analysis.

\subsection{The parametrics of higher order corrections to a stack of $\boldmath{\overline{D3}}$-branes} \label{sec:parametrics}

Before calculating higher order corrections to the potential of a stack of $\overline{D3}$-branes it is instructive to understand the parametrics of the setup. 

As summarized in Sec.~\ref{sec:kpv}, the NS5-brane potential including $\alpha'$ corrections enjoys an expansion in $g_sM$ and $p/M$. The parameter $g_sM$ comes from higher curvature and $F_3$/$H_7$ corrections which only see the $S^3$ at the tip of the throat. The $p/M$ expansion on the other hand is introduced by the higher order $F_2$ corrections which are sensitive to the size of the $S^2$ wrapped by the NS5-brane. Additionally, there are $C_2$ and extrinsic curvature corrections which are also controlled by the size of the $S^2$. 

From the $\overline{D3}$-brane perspective there are two types of corrections. On the one hand there are higher derivative corrections which are \textit{not} included in the nonabelian brane action \eqref{eq:dbi} and \eqref{eq:cs}. These are the same corrections that also exist on single branes. As the $\overline{D3}$-branes are pointlike in the internal space, these $\alpha'$ corrections do only see the size of the $S^3$ at the tip. Their expansion parameter is then $1/(g_sM)^2$.

On the other hand, there are higher commutator corrections when expanding the nonabelian action to higher powers in $\lambda$. Their parametrics is readily understood when looking at \eqref{eq:QKPV}. As we will see in Sect.~\ref{sec:commutatorcorr}, higher order commutator corrections are $\lambda^2\sim \alpha'^2$ suppressed compared to the leading order terms. From \eqref{eq:QKPV} we see that each $\lambda$ will either come with $[\Phi^i,\Phi^j]/g_s$ or with $g_s \Phi^l F_{lkj}$. Together with $\Phi^i\sim A\alpha^i \sim g_s^2f \alpha^i$ and \eqref{eq:f3nonabelian}, the commutator corrections will scale like 
\begin{equation}
    \lambda^2 g_s^6 f^4 (\alpha^i)^2 \sim p\frac{p^2-1}{M^2} \sim p\frac{p^2}{M^2} \,,
\end{equation}
where we used that $\tr((\alpha^i)^2)=p(p^2-1)$ and expanded for large $p$ in the last step. The overall factor of $p$ is common to every term in the potential. Hence, the commutator corrections are suppressed by $(p/M)^2$. This is expected since commutator corrections correspond to $F_2$ corrections from the dual abelian perspective.

\subsection{ $\boldmath{\alpha'}$ corrections to (anti)-D3-brane stacks} \label{sec:alphacorr}

Additional to the nonabelian action \eqref{eq:dbi} and \eqref{eq:cs} a stack of branes receives $\alpha'$ corrections in the same way as single branes do. The $\alpha'$ corrected action for a stack of $\overline{D3}$-branes then reads (see e.g.~\cite{Bena:2019rth} for a reference working with $\alpha'$ corrections to the nonabelian brane action)\footnote{We again neglect $F_2$ and all terms in the DBI part contributing to the kinetic terms of $\Phi^i$.}
\begin{equation}
\label{eq:antid3corrected}
    S = -\frac{T_3}{g_s} \int \dd^4\sigma\,\text{STr}\left( \sqrt{\det\left( G_{ab}
    \right)\det(Q^i_{~j}) } \left(1+ \alpha'^2\mathcal{L}_{\alpha'^2}\right)\right)
      + S_\text{CS} +S_{\text{CS},\alpha'^2}\,,
\end{equation}
where $S_{\text{CS},\alpha'^2}$ denotes $\alpha'$ corrections to the CS action and $\mathcal{L}_{\alpha'^2}$ denotes $\alpha'$ corrections to the DBI action. 

We relegate all the details to App.~\ref{sec:app} and only state the results here.
The non-vanishing terms in our setup can be found in \cite{Bachas:1999um,Garousi:2010ki,Robbins:2014ara,Garousi:2014oya} and are schematically written as (suppressing the index structure)
\begin{align}
\label{eq:alphacorrdbi}
    \mathcal{L}_{\alpha'^2} \sim & - R^2 + H_3^4 +F_3^4 + R H_3^2\\
    S_{\text{CS},\alpha'^2} \sim & \,\alpha'^2 \int\dd^4\sigma\,\Str\left( \varepsilon_{(4)} \nabla F_5 R\right)\,.
    \label{eq:alphacorrcs}
\end{align}
Two comments are in order.
First, note that the term $F_3^4$ in the DBI action is strictly speaking not derived (yet). We inferred their structure by the invariance of the $D3$-brane under S-duality at higher orders in $\alpha'$ as proposed in \cite{Green:1996qg,Bachas:1999um,Basu:2008gt,Garousi:2011fc}. Second, including $\alpha'$ corrections to the action of $\overline{D3}$-branes has the following advantage compared to including them into the NS5-brane action as done in \cite{Hebecker:2022zme,Schreyer:2022len}. Most of the corrections on $\overline{D3}$-branes are known and not need to be inferred by S-duality arguments. 

Referring to the App.~\ref{sec:app} for details, \eqref{eq:alphacorrdbi} and \eqref{eq:alphacorrcs} can be evaluated at the tip of the KS throat with the result
\begin{equation}
    \mathcal{L}_{\alpha'^2} =- \frac{c_1}{(g_sM)^2}\,,\qquad\qquad \text{and} \qquad\qquad S_{\text{CS},\alpha'^2} =- \frac{T_3 p }{g_s }\int\dd^4 x \sqrt{-g_4} \frac{c_2}{(g_sM)^2}\,,
    \label{eq:alphacorr}
\end{equation}
where $c_1 = 4.9059$ (see \eqref{eq:dbinonvanishingevaluated}) and $c_2 = 31.5953$ (see below \eqref{eq:csnonvanishingevaluated}). Note that the correction to the CS action will only induce a shift to the constant term of the KPV potential \eqref{eq:vonshellleading} and has no dependence on the noncommuting scalars $\Phi^i$. For the correction to the DBI action this is not true as they are multiplied by $\sqrt{\det(Q)^i_{~j}}$.

\subsection{Higher order commutator corrections on $\boldmath{\overline{D3}}$-brane stacks} \label{sec:commutatorcorr}

In this section, we calculate all commutator corrections arising from \eqref{eq:antid3} at $\mathcal{O}(\lambda^4)$. These are the leading, non-vanishing terms in the commutator expansion after the terms at $\mathcal{O}(\lambda^2)$ calculated in Sect.~\ref{sec:leadingorder}. All terms at $\mathcal{O}(\lambda^3)$ vanish under the maximally symmetric trace. This is a crucial property since after T-duality such $\lambda^3$ terms would correspond to terms cubic in $F_{ab}$ which are known to be absent.

Importantly, it should be emphasized that the leading order KPV potential from the NS5-brane perspective already takes into account all higher commutator corrections. The reason is that commutator corrections map to $F_2$ corrections on the abelian side which are already summed up in the square root of the DBI part of the abelian action.

We start by calculating commutator corrections to the DBI part of \eqref{eq:antid3} which can be done by expanding $\Str\sqrt{\det Q^i_{~j}}$ up to $\mathcal{O}(\lambda^4)$.

To do so, we define the $\Phi$-dependent terms of \eqref{eq:QKPV} as a $(3\times3)$-matrix $M$, i.e.~$Q=1+M$, and write\footnote{The following equation is written for a ($3\times3$)-matrix such that no terms of the form $\tr M^4$ can occur.}
\begin{equation}
\label{eq:expansion}
    \begin{split}
    \hspace*{-.4cm}&\sqrt{\det(1+M)}= 1+\frac{1}{2} \tr M 
        + \frac{1}{8}\left[(\tr M)^2 - 2\tr M^2 \right]
        +\frac{1}{48} \left[ (\tr M)^3 -6\tr M\tr M^2 + 8 \tr M^3\right]\hspace*{-.4cm}\\
       \hspace*{-.4cm} &\qquad\quad +\frac{1}{384} \left[ -7 (\tr M)^4 +36 (\tr M)^2 \tr M^2 -12 (\tr M^2)^2 - 32 \tr M \tr M^3 \right]
        +\mathcal{O}(M^5)~,\hspace*{-.4cm}
\end{split}
\end{equation}
where the trace is over the $i,j,k,\dots$ indices and not over the matrix indices $A,B,\dots$ of $(\Phi^i)^A_{~B}$.
The scalar potential from the DBI action up to $\mathcal{O}(\lambda^4)$ is then of the form
\begin{equation}
\label{eq:Vdbicorr}
    \begin{split}
        \hspace*{-.4cm}&V_{\text{DBI},\mathcal{O}(\lambda^4)} =  \frac{T_3}{g_s} \Biggl( \tr(\mathbbm{1})+\frac{\lambda^2}{4g_s^2}\tr\left( [\Phi^i,\Phi^j][\Phi^j,\Phi^i] \right) - \frac{i\lambda^2}{6} F_{ikl} \tr\left([\Phi^i,\Phi^k]\Phi^l\right)\hspace*{-.4cm} \\
        \hspace*{-.4cm}&\quad-\frac{\lambda^4}{72}\Str\left([\Phi^i,\Phi^k]\Phi^l [\Phi^j,\Phi^m]\Phi^p\right) F_{lki}F_{pmj}+\frac{\lambda^4}{36} \Str\left([\Phi^i,\Phi^k]\Phi^l[\Phi^j,\Phi^m]\Phi^p\right)F_{lkj} F_{pmi} \hspace*{-.4cm}\\
       \hspace*{-.4cm} &\quad+ \frac{i\lambda^4}{24g_s^2}\Str\left([\Phi^i,\Phi^k]\Phi^l[\Phi^j,\Phi^m][\Phi^m,\Phi^j]\right)F_{lki} -\frac{\lambda^4}{32g_s^4}\Str\left(\left([\Phi^i,\Phi^j][\Phi^j,\Phi^i]\right)^2\right) \hspace*{-.4cm}\\
        \hspace*{-.4cm}&\quad-\frac{i\lambda^4}{6g_s^2}\Str\left([\Phi^i,\Phi^j][\Phi^j,\Phi^m][\Phi^m,\Phi^k]\Phi^l F_{lki}\right)\Biggr)\,,\hspace*{-.4cm}
    \end{split}
\end{equation}
where we have already used the symmetries of the maximally symmetric trace. In particular, all terms at $\mathcal{O}(\lambda^3)$ vanish upon taking the STr. 

We proceed with the CS action where we additionally have to include the nonabelian features 2) and 3) summarized in Sec.~\ref{sec:actions}. 
The only contribution at $\mathcal{O}(\lambda^4)$ comes from the term
\begin{equation}
    \frac{T_3 \lambda^2 }{4} \int \Str \left(P\left[(\iphi\iphi)^2 C_2\wedge C_2\wedge C_4\right]\right) = \frac{T_3 \lambda^2 }{2} \int \Str \left(P\left[(\iphi\iphi C_2)^2 C_4\right]\right)\,,
\end{equation}
where we used that $C_4$ only has indices along the brane stack. After evaluating the pullback and nonabelian Taylor expanding $C_2$, one finds a contribution to the potential of the form
\begin{equation}
\label{eq:Vcscorr}
    \frac{g_s \,V_\text{CS}}{T_3} \supset  - \frac{\lambda^4}{72} \Str \left([\Phi^i,\Phi^j]\Phi^k[\Phi^l,\Phi^m]\Phi^n\right) F_{kji} F_{nml}\,,
\end{equation}
which equals the first term in the second line in \eqref{eq:Vdbicorr} as one would have expected due to the imaginary self-dual flux background.\footnote{The second term in the second line of \eqref{eq:Vdbicorr} cannot be obtained from any term in the CS action since due to the nonabelian interior product in the CS action each $F_{ijk}$ will always be contracted with a commutator $[\Phi^i,\Phi^j]$.}
For completeness, the total contribution from the CS part of the action \eqref{eq:antid3} up to order $\lambda^4$ is given by
\begin{equation}
\label{eq:}
    \frac{g_s\,V_{\text{CS},\mathcal{O}(\lambda^4)}}{T_3} = -\frac{i\lambda^2}{6} F_{ikl} \,\tr\left([\Phi^i,\Phi^k]\Phi^l\right)- \frac{\lambda^4}{72} \Str \left([\Phi^i,\Phi^j]\Phi^k[\Phi^l,\Phi^m]\Phi^n\right) F_{kji} F_{nml}\,.
\end{equation}

\subsection{The higher order result}

In this section, we go through the same steps as in Sect.~\ref{sec:leadingorder} but use instead the $\alpha'$ corrected scalar potential which includes all corrections from Sects.~\ref{sec:alphacorr} and \ref{sec:commutatorcorr}. We calculate the corrected stationary points of the potential, the corrected radius of the fuzzy $S^2$, and evaluate the potential at the minimum.

We obtain the $\alpha'$ corrected potential at order $\lambda^4$ by collecting all terms in the scalar potential derived in Sect.~\ref{sec:alphacorr} and \ref{sec:commutatorcorr}:
\begin{equation}
\label{eq:Vcorr}
    V_\text{tot} = V_{\text{DBI},\mathcal{O}(\lambda^4)}\left(1 - \frac{c_1}{(g_sM)^2}\right) + V_{\text{CS},\mathcal{O}(\lambda^4)} +\frac{T_3\,p}{g_s}\, \frac{c_2}{(g_sM)^2}\,,
\end{equation}
where $c_{1,2}$ are due to $\alpha'$ corrections of Sect.~\ref{sec:alphacorr} and are given by \eqref{eq:alphacorr}, $V_{\text{DBI},\mathcal{O}(\lambda^4)}$ is from \eqref{eq:Vdbicorr}, and $V_{\text{CS},\mathcal{O}(\lambda^4)}$ from \eqref{eq:Vcscorr}.

The stationary point of the scalar potential $V_{\mathcal{O}(\lambda^4)}$ determined by
\begin{equation}
    \frac{\delta V_{\mathcal{O}(\lambda^4)}}{\delta \Phi^n } = 0\,,
    \label{eq:eom}
\end{equation}
which we solve as in Sect.~\ref{sec:leadingorder} with the ansatz $[\Phi^i,\Phi^j] = A \,\varepsilon_{ijk} \Phi^k$, where $\Phi^i= -i A \alpha^k/2$.
Plugging in the ansatz into \eqref{eq:eom} leads after a long but straightforward calculation to 
\begin{equation}
\label{eq:eomcompact}
    0 = \left( \lambda^2 A^2 \left(\frac{i b A}{g_s^2}-\frac{f}{2}(b+1) \right)-i\lambda^4A^5 \left(\frac{f^2}{9} + \frac{7ib A f}{18 g_s^2} +\frac{bA^2}{3g_s^2}\right)\left(\frac{3}{8} C -\frac{1}{2}\right)\right) \alpha^n\,,
\end{equation}
where we abbreviated $b=1-c_1/(g_sM)^2$ and $C=p^2-1$ denotes the quadratic Casimir.

Eq.~\eqref{eq:eomcompact} can be solved perturbatively for $A$ with the ansatz
\begin{equation}
    A = -i g_s^2 f \gamma (1+\delta)\,,\qquad \qquad \gamma =\frac{b+1}{2b}\,.
    \label{eq:ansatzcorr}
\end{equation}
This ansatz can be understood as follows: The leading order solution of Sect.~\ref{sec:leadingorder} is $A=-ig_s^2f$ and $\gamma$ takes into account the $\alpha'$ corrections at order $\lambda^2$ in the commutators. So neglecting the commutator corrections at $\mathcal{O}(\lambda^4)$, the solution would be $A=-ig_s^2 f \gamma$. Expanded for large $g_sM$, $\gamma$ is given by
\begin{equation}
    \gamma = 1 + \frac{c_1}{2(g_sM)^2} + \frac{c_1^2}{2(g_sM)^4}+\cdots\,,
\end{equation}
such that the leading order solution is recovered in the limit $g_sM\to\infty$. Finally, the correction $\delta$ comes from $\alpha'$ corrected commutator terms at order $\lambda^4$. Since we solve \eqref{eq:eomcompact} perturbatively, our solution will be valid in the regime where $\delta\ll1$.
 
Inserting the ansatz \eqref{eq:ansatzcorr} into \eqref{eq:eomcompact} and neglecting all terms suppressed at least by $\delta^2$ compared to the leading order term\footnote{This is allowed since we only aim to find the leading order term in the correction $\delta$. For calculating higher order terms, the potential needs to be expanded to higher order in $\lambda$.}, one finds
\begin{equation}
\label{eq:eomsol}
\begin{split}
    \delta &= \lambda^2 f^4 g_s^6 \gamma^2\left(\frac{3}{8}C-\frac{1}{2} \right) \left(\frac{\gamma^2}{3}-\frac{7\gamma}{18}-\frac{1}{9b}\right) + \mathcal{O}(\lambda^4 f^8 g_s^{12})\\
    & = \lambda^2 f^4 g_s^6 \gamma^2\left(\frac{3}{8}C-\frac{1}{2} \right)\left(- \frac{1}{6}+\frac{ c_1}{36(g_sM)^2}+ \mathcal{O}((g_sM)^{-4})\right) + \mathcal{O}(\lambda^4 f^8 g_s^{12})\,.
    \end{split}
\end{equation}
As expected from Sect.~\ref{sec:parametrics}, the higher commutator corrections are suppressed by $\lambda^2g_s^6f^4C\sim(p/M)^2$.
Importantly, the leading order term in the expression for $\delta$ is negative for $p\geq2$ and $\alpha'$ corrections contribute positively to $\delta$. This will be crucial when determining the corrected radius of the fuzzy $S_2$ below. 

Before doing so, we calculate the on-shell value of the effective potential using the ansatz~\eqref{eq:ansatzcorr}. We find 
\begin{equation}
\label{eq:Vonshell}
    \begin{split}
        V_{\text{tot}} =& \frac{T_3 \,p}{g_s} \Biggl( 1 + \frac{c_2-c_1}{(g_sM)^2}+ \lambda^2g_s^6f^4(p^2-1)\left(\frac{b\,\gamma(1+\delta)}{8}-\frac{b+1}{12}\right)(1+\delta)^3\gamma^3 \\
        &- \lambda^4 f^8 g_s^{12} \left(C^2-\frac{4}{3}C\right) \frac{\gamma^6 (1+\delta)^6}{16}\left(-\frac{1}{18}-\frac{\gamma\,(1+\delta)b}{6} + \frac{\gamma^2(1+\delta)^2b}{8}\right)\Biggr)\,,
    \end{split}
\end{equation}
where we used that $\Str(\alpha^i\alpha^i\alpha^j\alpha^j)=C^2-4C/3$. One can convince oneself that the energy of the puffed up brane stack is positive everywhere except close to $g_sM\sim \sqrt{c1}$ and smaller than the energy of the stack when all branes coincide, i.e.~$A=0$. The puffed up state is therefore energetically favored.

Next, we calculate the corrected radius $R_{S^2}$ of the fuzzy $S^2$. As reviewed in Sec.~\ref{sec:leadingorder}, the radius is given by 
\begin{equation}
\label{eq:rs2fuzzycorr}
    R_{S^2}^2 \simeq \frac{\lambda^2}{p } \tr\left((\Phi^i)^2\right) = \lambda^2 g_s^4 f^2(p^2-1) \frac{\gamma^2(1+\delta)^2}{4}\,,
\end{equation}
where we used \eqref{eq:ansatzcorr}. Plugging in $\delta$ from \eqref{eq:eomsol}, $\gamma$, and $f$ yields for $p\gg1$ (the result for general $p$ is obtained upon replacing $p$ by $\sqrt{p^2-1}$)
\begin{equation}
    R_{S^2} \simeq \frac{\lambda \,p}{b_0^4 M} R_{S^3} \left(1 - \frac{\lambda^2}{b_0^{12}}\left(\frac{p}{M}\right)^2 + \frac{c_1 \left(3b_0^{12}-8(p/M)^2\lambda^2\right)}{6b_0^{12}(g_sM)^2}\left(1+\frac{c_1}{(g_sM)^2}\right) + \cdots\right)\,.
    \label{eq:rs2corrected}
\end{equation}

As a result, $\alpha'$ and commutator corrections have a different impact on the radius of the puffed up brane configuration. The commutator corrections (suppressed by $(p/M)^2$) facilitate the existence of the classical minimum because they make $R_{S^2}$ smaller\footnote{This interpretation is of course only true as long as the commutator corrections are small, the commutator expansion controlled, and our solution hence still valid.} whereas $\alpha'$ curvature corrections (suppressed by $1/(g_sM)^2$) complicate the existence of the minimum as they enlarge $R_{S^2}$. A larger value of $R_{S^2}$ endangers the metastable state of the brane configuration since the minimum is driven closer to the equator of the $S^3$ at the tip of the throat where the $\overline{D3}$-branes start to annihilate with flux to form a supersymmetric state at the north pole of the $S^3$.

\subsection{Conditions for control and phenomenological implications}\label{sec:pheno}

Before analyzing the implications for phenomenology, we discuss the crucial question of where in the $(g_sM,p/M)$ parameter space our calculations of the preceding sections are trustworthy. 

In order for the flat space approximation in \eqref{eq:Gnonabelian} and \eqref{eq:Cnonabelian} to be valid and to control the $\alpha'$ corrections of Sect.~\ref{sec:alphacorr}, $g_sM$ should be sufficiently large compared to the constant\footnote{Note that we do not require $(g_sM)^2$ to be sufficiently large compared to $c_2$ since $c_2$ increases the effective tension of the brane where instead the corrections related to $c_1$ can render the brane tachyonic.} $c_1$, i.e.
\begin{equation}
\label{eq:controlalpha}
(g_sM)^2>c_1\,. 
\end{equation}
Otherwise, the higher order terms compete with the tree level contribution signaling a breakdown of the $\alpha'$ expansion. Additionally, the potential in the minimum will turn negative around $(g_sM)^2\approx c_1$. In the following we therefore stick to the regime \eqref{eq:controlalpha}.

Moreover, the solution $\delta$ of \eqref{eq:eomsol} is only controlled as long as $|\delta|\ll1$ such that when solving the equation of motion \eqref{eq:eomcompact} for $\delta$, terms of order $\delta^2$ and higher can be neglected. 
This translates into the constraint
\begin{equation}
    |\delta| = \frac{\lambda^2}{b_0^{12}}\frac{p^2-1}{M^2} \left(1+ \frac{5c_1}{6(g_sM)^2}\right)+\cdots \ll 1\,,
    \label{eq:deltaconstraint}
\end{equation}
which requires $p/M$ to be small enough.

More importantly, we can only trust the scalar potential and the expansion in commutators when the Taylor expansion of the square root in the DBI action converges rapidly~\cite{Myers:1999ps}.\footnote{Equivalently, this requires the typical distance between the branes inside the brane stack to be much smaller than the string length~\cite{Myers:1999ps}. From the area element $4\pi R_{S^2}^2/(g_sp)$ of the fuzzy $S^2$, this is fulfilled if $R_{S^2}\ll \sqrt{g_s \,p}\, l_s$. The factor of $g_s$ appears since we are working in the S-dual frame.}
Commutator terms are always suppressed by $\lambda^2$ and each $\lambda$ either comes with $[\Phi^i,\Phi^j]/g_s$ or $g_s\Phi^kF_{ijk}/3$. Using both expressions together with \eqref{eq:ansatzcorr} and $\Phi^i\sim A \alpha^i/2$, higher commutator terms are suppressed if
\begin{equation}
\label{eq:controlpotential}
    f_p\equiv\frac{4\lambda^2}{b_0^{12}} \frac{p^2-1}{M^2} \gamma^2 (1+\delta)^2 \max\left[1/9, \gamma(1+\delta)/3,\gamma^2(1+\delta)^2\right] \ll 1\,,
\end{equation}
where the three different terms arise from all possible contributions at order $\lambda^2$.

Finally, as discussed already in Sect.~\ref{sec:leadingorder}, our setup of the puffed up brane stack can only be classically stable if the radius of the fuzzy $S^2$ is much smaller than the radius of the $S^3$. For large $p$, this translates into (see also \eqref{eq:rs2corrected} for the expansion in $p/M$ and $g_sM$)
\begin{equation}
\label{eq:controlminimum}
    \frac{R_{S^2}}{R_{S^3}} = \frac{\lambda}{b_0^4} \frac{p}{M} \gamma \,(1+\delta) \equiv c_m \ll 1\,.
\end{equation}
In the limit $\gamma\to1$ together with $\delta\to0$, we recover \eqref{eq:treelevelconstraint}.
In the major part of the $(g_sM,p/M)$ parameter space, the control over the minimum of the potential will turn out to be the strongest constraint since it is neglecting $\mathcal{O}(1)$ prefactors the square root of \eqref{eq:controlpotential}. For larger values of $p/M$ and small $g_sM$, the control over the potential is lost slightly faster than the control over the minimum. 

\begin{figure}[h]
    \centering
    \includegraphics[width=.8\textwidth]{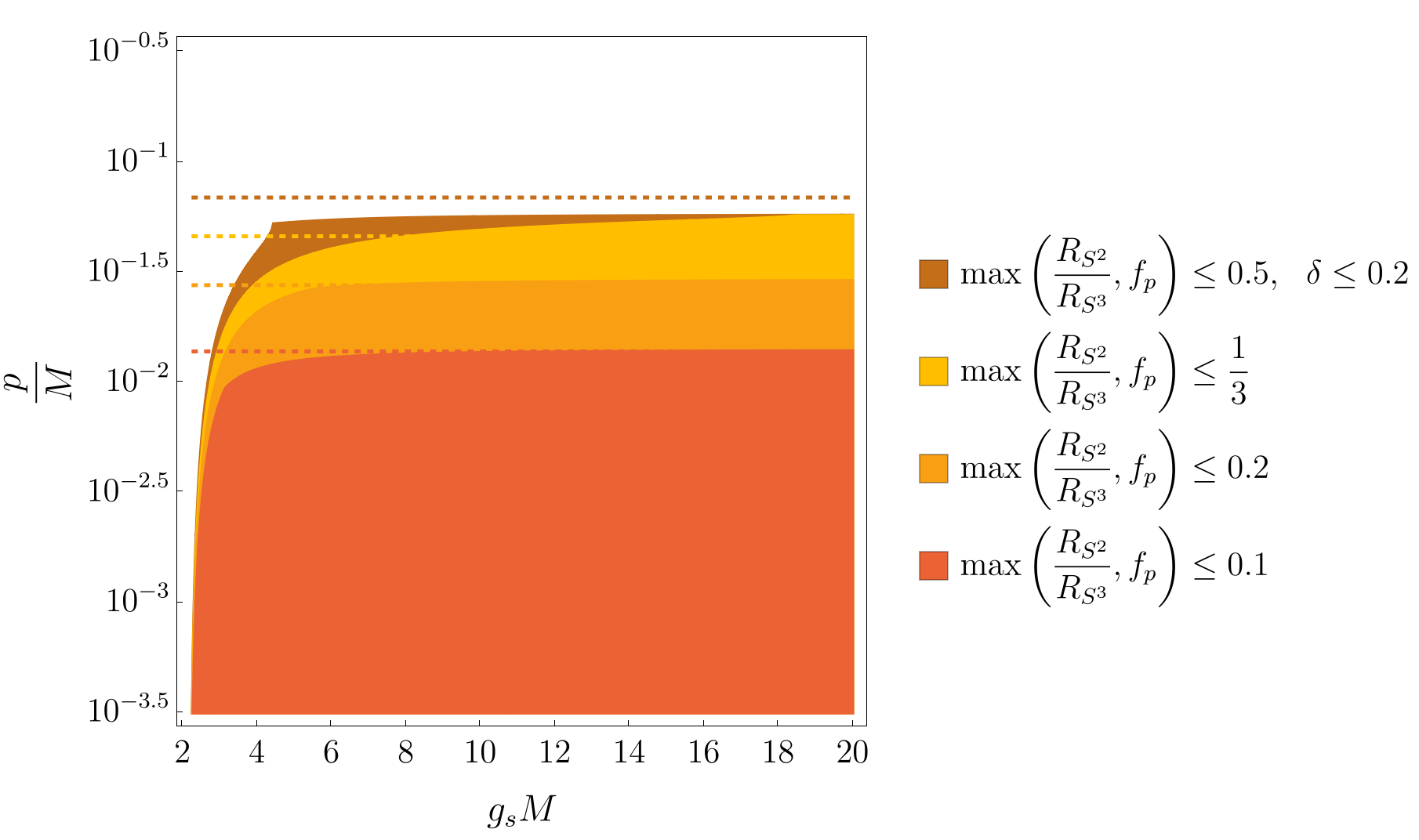}
    \caption{
    Plot of the control conditions \eqref{eq:deltaconstraint} - \eqref{eq:controlminimum} for different values of $c_m$. Equ.~\eqref{eq:controlminimum} is the necessary condition that a metastable minimum exists. The dotted lines correspond to the tree level version of the condition for a minimum to exist (see \eqref{eq:treecond}).
    Note that the region where $c_m<0.5$ is bounded by the constraint \eqref{eq:deltaconstraint}, i.e.~$|\delta|<0.2$ for $g_sM>4$.}
    \label{fig:control}
\end{figure}

Phenomenologically, the constraint \eqref{eq:controlminimum} is of course very relevant since it describes a necessary condition for the metastable minimum to exist and hence also for the $\overline{D3}$-brane uplift to work. It is therefore instructive to plot regions in the $(g_sM,p/M)$ parameter space where $c_m$ is smaller than a given value describing the reader's notion of control by still retaining control over the potential \eqref{eq:controlpotential} and the solution of the equation of motion \eqref{eq:deltaconstraint}. 
In Fig.~\ref{fig:control}, we therefore plot 
\begin{equation}
\label{eq:constraint}
    \max(R_{S^2}/R_{S^3},f_p) \leq c_m
\end{equation}
for some example values of $c_m$. Additionally, we require $\delta \leq \delta_\ast =0.2$ for having control over the solution.\footnote{The choice of $\delta_\ast$ is clearly up to the reader. It is important to note that the constraint \eqref{eq:deltaconstraint} (independently of the choice of $\delta_\ast$) only becomes important for values $c_m\geq 1/3$ where we start to distrust the potential and the existence of a classically stable minimum anyways.} Hence, the colored regions in Fig.~\ref{fig:control} ensure an order of control given by $c_m$ over the potential and the existence of the minimum.

The dotted lines in Fig.~\ref{fig:control} correspond to the tree level constraint for a minimum to exist given an order of control determined by $c_m$. The condition reads for large $p$ \eqref{eq:treelevelconstraint}
\begin{equation}
    \left(\frac{R_{S^2}}{R_{S^3}}\right)_\text{l.o.~KPV} = \frac{p}{M} \frac{\lambda}{b_0^4} \equiv c_m \ll 1\,.
    \label{eq:treecond}
\end{equation}
In Fig.~\ref{fig:control}, one can see that in the phenomenologically interesting regime $g_sM\lesssim6$, the tree level and $\alpha'$ corrected constraints differ considerably from each other since $\alpha'$ corrections start to become relevant. This shows that it is crucial to take these corrections into account when searching for the minimal value of $g_sM^2$.
Furthermore, for large $g_sM$ and small $p/M$, the $\alpha'$ corrected constraints asymptote towards the tree level constraint, as it should be. Increasing $p/M$ at large $g_sM$, the $\alpha'$ corrected constraints become weaker than the tree level version (see the dotted line for $c_m=1/3$). This is expected since the commutator corrections facilitate the existence of the minimum\footnote{Again, this is only true as long as the commutator corrections are small and one trusts the solution.} as they decrease $R_{S^2}$. If one increases $p/M$ even further, one loses control over the commutator expansion and the tree level and $\alpha'$ corrected constraints cannot be compared anymore (see brown region and dotted line for $c_m=0.5$).

\begin{figure}
    \centering
    \includegraphics[width=.8\textwidth]{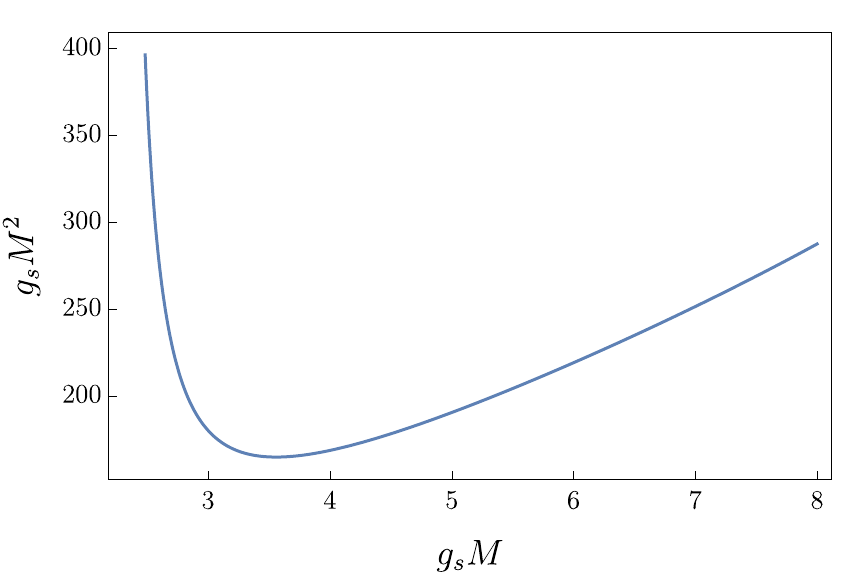}
    \caption{The value of $g_sM^2$ as a function of $g_sM$ for the value $c_m=0.2$. We choose $p=1$ for convenience.}
    \label{fig:gsm202}
\end{figure}

The constraint \eqref{eq:constraint} can also be translated into a bound on $g_sM^2$ by calculating $g_sM^2$ for different values of $c_m$ on the boundary curves of the colored regions in Fig.~\ref{fig:control}. The boundary curves with $c_m\leq 1/3$ (where the constraint \eqref{eq:deltaconstraint} does not play a role) can be fitted nicely with a Fernández-Guasti squircle\footnote{The author thanks Maximilian Peter Wollner for pointing this out.} \cite{guasti2005lcd}
\begin{equation}
    \log _{10}\left[\frac{p}{M}\right] = 
    \begin{cases}
    \left(\frac{p}{M}\right)_0 + \alpha \sqrt{\frac{1-\left(\frac{g_sM-(g_sM)_0}{\beta}\right)^2}{1-s\left(\frac{g_sM-(g_sM)_0}{\beta}\right)^2}}, & 2.25 <  g_sM\leq(g_sM)_\ast\,,\\
    \left(\frac{p}{M}\right)_\infty, & g_sM\geq (g_sM)_\ast\,.
    \end{cases}
    \label{eq:fit}
\end{equation}
\begin{table}[!h]\centering
	\caption{Fitting parameters for the boundary curves where $c_m\leq1/3$.}
	\vspace{.3cm}
	\label{tab:fit}
		\begin{tabular}{c||ccccccc}
		\toprule
		$c_m$ & $\left(\frac{p}{M}\right)_0$ & $\alpha$ & $(g_sM)_0$ & $\beta$ & $s$ & $ \left(\frac{p}{M}\right)_\infty$ & $(g_sM)_\ast$ \\
		\midrule
		0.1& -14.59&12.7 &5.07 &2.93 &0.98 & -1.895  & 5\\
       \midrule
       0.2&-4.18 & 2.63 & 7.48 & 5.22 & 0.93 & -1.556 & 8\\
       \midrule
       $\frac{1}{3}$ & -4.01 & 2.72 & 22.16 & 19.9 & 0.98& -1.317 & 8\\
       \bottomrule
	\end{tabular}
\end{table}
The fitting parameters are given in Tab.~\ref{tab:fit}. Using \eqref{eq:fit} we can calculate $g_sM^2$ on the boundary curve by fixing the number of branes $p$. This is depicted in Fig.~\ref{fig:gsm202} where the dependence of $g_sM^2$ on $g_sM$ for $p=1$ and $c_m=0.2$ is shown.\footnote{As we will also comment on in Sect.~\ref{sec:comparison}, one has to choose $p>1$ for the brane stack to be nonabelian. For $p=1$, the brane does not puff up. If we choose $p=1$ we only do so in order to compare the bounds to the known bounds in the literature \eqref{eq:gm2leading} and \eqref{eq:boundalpha}.} From there, we can read off the minimal value of $g_sM^2$. The results are shown in Tab.~\ref{tab:gsm2}.
\begin{table}[!h]\centering
	\caption{Minimal values of $g_sM^2$, their corresponding values of $g_sM$ and $p/M$ for different values of $c_m$.}
	\vspace{.3cm}
	\label{tab:gsm2}
		\begin{tabular}{c||ccc}
		\toprule
		$c_m$ & $\left(g_sM^2\right)_\text{min}$ & $\left(g_sM\right)_\text{min}$ & $\left(\frac{p}{M}\right)_\text{min}$ \\
		\midrule
		0.1& 302$\times p$ & 3.23& 0.0107\\
       \midrule
       0.2&165$\times p$ & 3.559 & 0.0216\\
       \midrule
       $\frac{1}{3}$ & 106$\times p$ & 3.714 & 0.0224\\
       \midrule
       0.5 & 83$\times p$ & 4.369 & 0.0532 \\
       \bottomrule
	\end{tabular}
\end{table}
The minimal value of $g_sM^2$ strongly depends on the choice of $c_m$ and is realized, as expected, at small values of $g_sM$ where $\alpha'$ corrections already have a significant impact. Interestingly, performing the above analysis taking into account only one of the constraints \eqref{eq:controlpotential} and \eqref{eq:controlminimum}, one finds that the minimal bounds on $g_sM^2$ change only marginally. So even if one is willing to attenuate the control of one of the constraints, the final bound on $g_sM^2$ stays similarly strong. 

The main limiting factor for achieving lower minimal values of $g_sM^2$ from the $\overline{D3}$-brane perspective is the required smallness of $p/M$ in order to control the expansion in commutators. Additionally, $p/M$ determines $R_{S^2}$ in relation to $R_{S^3}$ (see e.g.~\eqref{eq:rfuzzy}) and therefore drives the control over the minimum.

Further, we can ask the question whether the new uplifting mechanism proposed in \cite{Hebecker:2022zme,Schreyer:2022len} can be verified from the $\overline{D3}$-brane perspective. 

The new uplifting mechanism makes use of the fact that the potential in the minimum can get negative and hence can be tuned arbitrarily close to zero energies (described by the curve \eqref{eq:zeroenergy} in $(g_sM,p/M)$ space) by tuning the parameters $g_sM$ and $p/M$ of the throat (cf.~Fig.~\ref{fig:kpv}). Hence, the uplifting energy can be tuned to be exponentially small without an exponentially large warp factor. The advantage is then that one side-steps all control issues related to long throats in KKLT \cite{Carta:2019rhx,Gao:2020xqh} and tadpoles in the LVS \cite{Gao:2022fdi} which are the main obstacles of a controlled $\overline{D3}$-brane uplift. The disadvantage is that the mechanism only works in the regime of parameter space where $\alpha'$ corrections and tree level terms are of approximately of same size. 

By evaluating the potential \eqref{eq:Vonshell}, it turns out that the potential is strictly positive everywhere except close to $g_sM = \sqrt{c_1}$ which is precisely at the very boundary of the control over $\alpha'$ corrections. In this regime, we face the same control issues that arise from the NS5-brane perspective and are therefore unable to access the regime $g_sM\leq 2.25$. 

This implies that only at small $g_sM$ where $\alpha'$ corrections are large the new uplifting mechanism still has a chance to exist.
In order to prove that the new uplifting mechanism actually works for small $g_sM$ requires taking into account higher order $\alpha'$ corrections which are currently not known.
On the other hand, for larger $g_sM$ and small $p/M$ the $\overline{D3}$-brane calculations are well controlled as can be seen from Fig.~\ref{fig:control} and the potential is strictly positive. We conclude that the new uplifting mechanism does not exist in this regime. 

One should note that the discussion of whether the new way of uplifting exists for large $g_sM$ is in some sense academic since the advantages of the new mechanism exist only for small $g_sM$ for which the throats are not too large.

\section{Limitations of the results and comparison with the NS5-brane perspective} \label{sec:comparison}

In this section we compare the results of \cite{Hebecker:2022zme,Schreyer:2022len} from the NS5-brane and the results of the previous sections from the $\overline{D3}$-brane perspective in more detail. We discuss how they complement each other and point out their limitations.

\begin{figure}[h]
    \centering
    \subfigure[]{\includegraphics[width=0.49\textwidth]{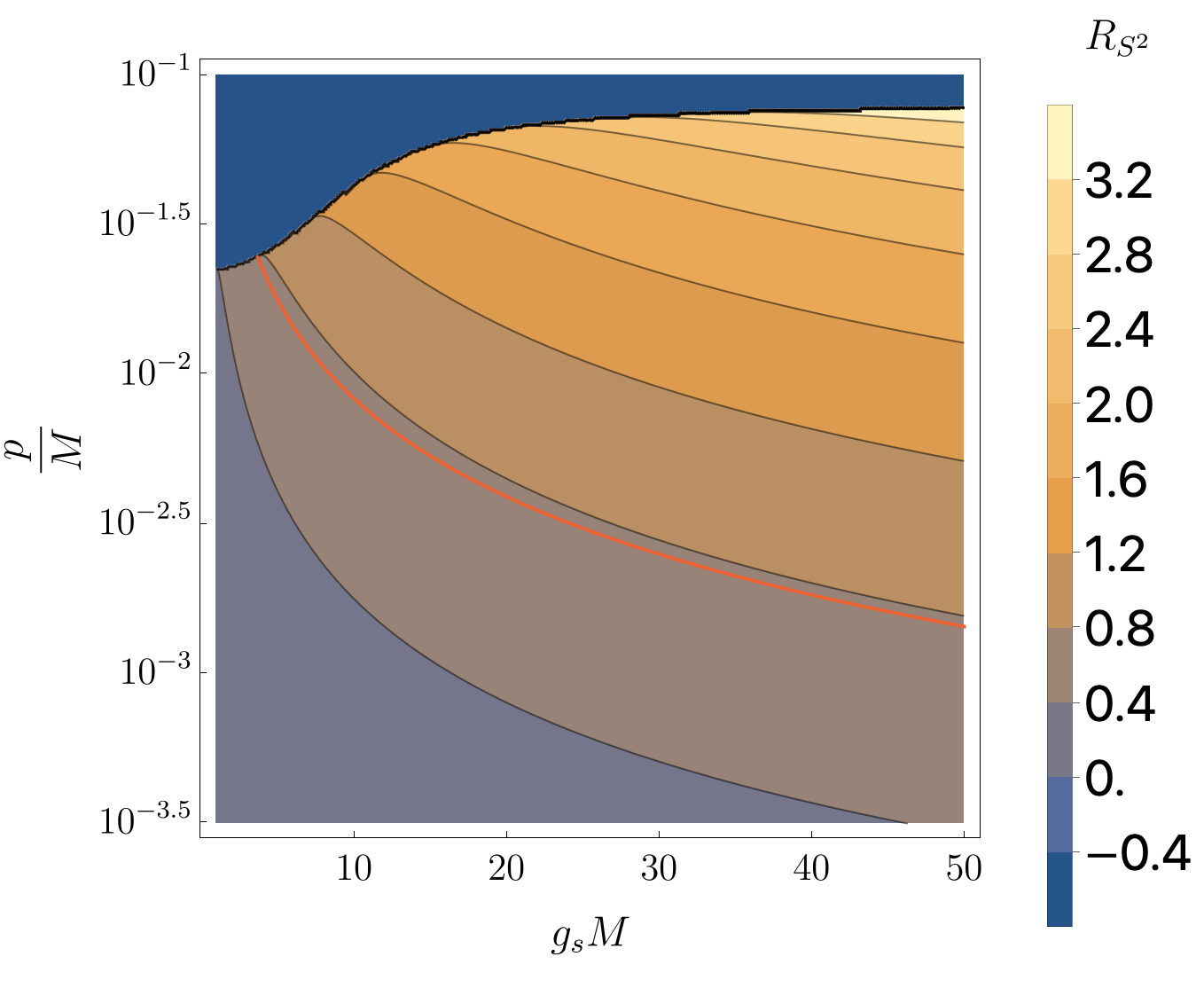}} 
    \subfigure[]{\includegraphics[width=0.49\textwidth]{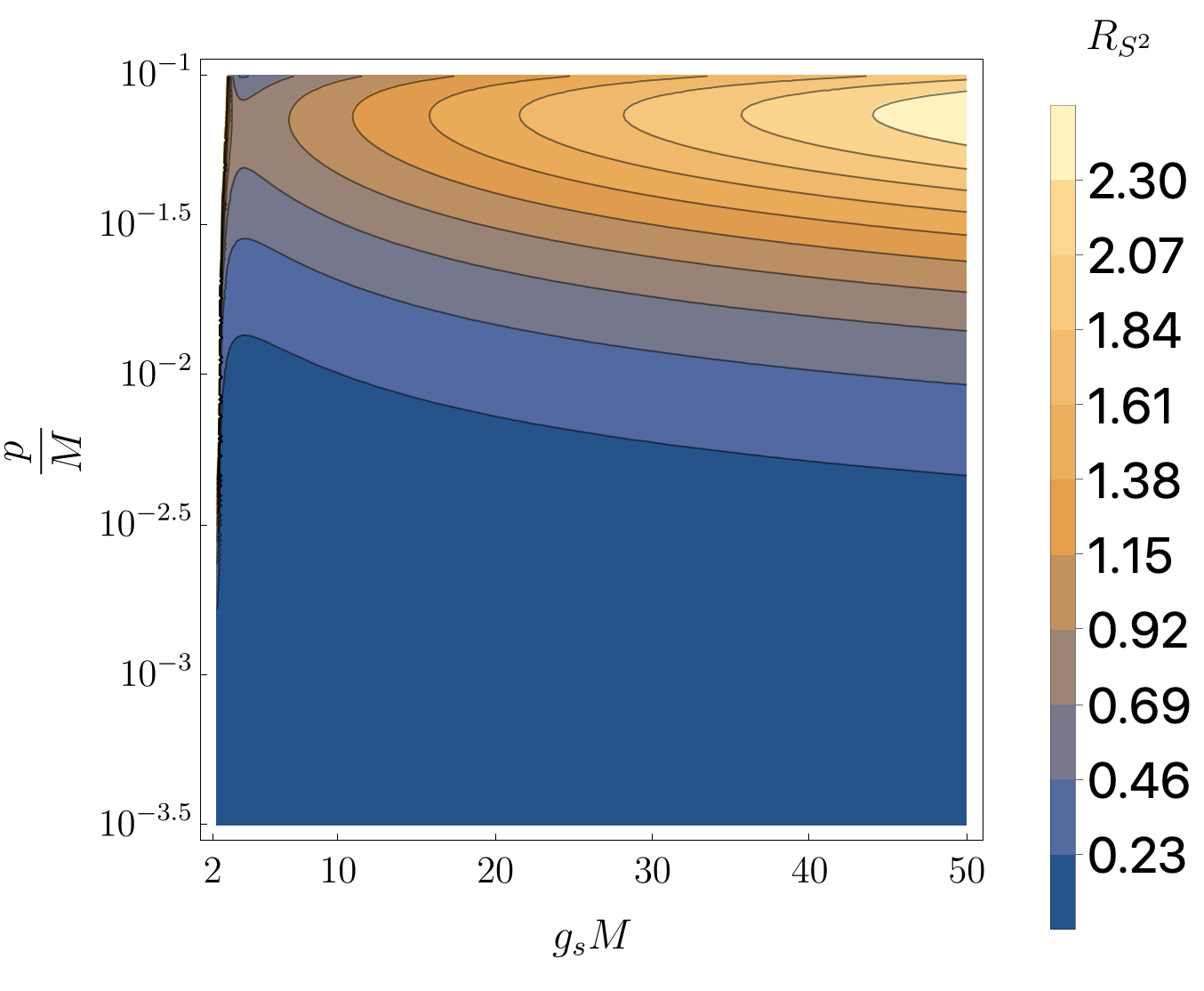}} 
    \caption{Comparison of the regions of control in the $(g_sM,p/M)$-plane from (a) the NS5-brane perspective and (b) the $\overline{D3}$-brane perspective. Fig.~(a) is adapted from \cite{Schreyer:2022len} and depicts the radius of the $S^2$ wrapped by the NS5-brane in the minimum of the potential. The red line indicates where the minimum is at zero energy. Fig.~(b) is a contour plot of the $R_{S^2}$ \eqref{eq:rs2fuzzycorr} obtained from the stack perspective. In both figures we have chosen $p=1$.}
    \label{fig:rs2}
\end{figure}

Most importantly, the regions of control of both pictures is different. The NS5-brane picture is valid for $R_{S^2}\gg l_s$ when $\alpha'$ corrections are controlled whereas the $\overline{D3}$-brane picture is valid for $R_{S^2}\ll \sqrt{g_s\,p}\,l_s$. Studying both perspectives therefore enables us to study the KPV setup in a broad region in parameter space. 
To visualize this, the radius $R_{S^2}$ is depicted in Fig.~\ref{fig:rs2} (a) for the NS5-brane and (b) for the stack of $\overline{D3}$-branes. 

As one can see from Fig.~\ref{fig:rs2}, the NS5-brane calculations are controlled for large $p/M$ and $g_sM$ where $R_{S^2}$ is large, and the $\overline{D3}$-brane calculations are controlled for small $p/M$ as can also be seen from Fig.~\ref{fig:control} since the expansion in commutators in only controlled for small $p/M$.
The value of $R_{S^2}$ does not match very precisely from both perspectives, however, a trend is clearly visible. We comment on the origin of this mismatch below.

Due to the different validity regimes each perspective is advantageous for answering different questions. The NS5-brane perspective is suitable for studying the classical decay into the SUSY minimum at $\psi=\pi$ since this requires control when $R_{S^2}\approx R_{S^3}$ with $R_{S^3} > l_s$. From the $\overline{D3}$-brane perspective the decay is not visible as we have assumed that the directions in which the brane stack expands into the fuzzy $S^2$ are flat, i.e.~$R_S^3\gg\ l_s$, such that the stack does not see the KS geometry. Therefore, we have assumed by hand that $R_{S^2}\ll R_{S^3}$ in order for the minimum to exist. To be able to see the KPV decay from the stack perspective one would need to extend the above analysis to the curved KS background and additionally choose $p$ large enough such that the stack perspective is controlled even for $R_{S^2}\approx R_{S^3}$.

The advantage of the $\overline{D3}$-brane picture is that calculations are controlled when $R_{S^2}$ is small. This is phenomenologically \textit{the} important regime since in this regime the bounds on $g_sM^2$ are the weakest. The $\overline{D3}$-brane picture is trustworthy for calculating the value of the potential in the minimum and the value of $R_{S^2}$ even if the $S^2$ is of string size.
We have seen an explicit example where interpreting the results from the NS5-brane perspective beyond the region of control led to wrong results:
The NS5-brane potential becomes negative when $R_{S^2}< l_s$ but the potential for the stack of $\overline{D3}$ showed that this does not happen -- the potential is positive everywhere except when $\alpha'$ corrections are more important than the tree level contribution, i.e.~$(g_sM)^2 \leq c_1 $. To study this regime, even higher order $\alpha'$ effects have to be taken into account.

It is also possible to compare the minimal bound on the quantity $g_sM^2$ from both perspectives. The NS5-brane analysis has led \cite{Schreyer:2022len} to a minimal bound of $g_sM>144$ occurring in the $(g_sM,p/M)$-plane at the very boundary of control, i.e.~at $R_{S^2}\approx l_s$. In this work, we are able to quantify the amount of control by the parameter $c_m$ with some example values given in Tab.~\ref{tab:gsm2}. 
Our analysis supports the results obtained from the NS5-brane perspective as we also find a bound of the form $g_sM^2\approx \mathcal{O}(100)$. The minimal value of $g_sM^2$ is therefore much larger than naively expected from leading order KPV. The precise numbers are marginally weaker from the stack perspective depending on the amount of control one wishes to have.

Before concluding, we discuss the limitations and open issues of our results. 
\begin{itemize}
    \item We have included only all known $\alpha'^2$ corrections to the $\overline{D3}$-brane action and by that neglected all other terms. This in particular includes couplings on the DBI action involving $\nabla F_5$ at order $\alpha'^2$. These terms can in principle be non-vanishing and contribute to the constant $c_1$. Their derivation would hence be crucial. Furthermore, to access the regime of small $g_sM$ would require to include even higher order $\alpha'$ corrections. Ideally, one would like to study the setup from the holographic picture which is perturbatively controlled for $g_sM\ll 1$. This could also shed light on the new way of uplifting in the small $g_sM$ regime and whether the potential in the minimum can actually be negative. 
    
    \item In our nonabelian calculations in Sects.~\ref{sec:leadingorder} and \ref{sec:commutatorcorr} we have assumed that the background in which the brane stack expands is flat, i.e.~that $R_{S^3}\gg l_s$, which is in principle a good approximation since we only control the stack perspective for $R_{S^2}\ll \sqrt{g_s\,p}\,l_s \leq l_s \ll R_{S^3}$. 
    Nevertheless, in the phenomenological relevant regime of smallish $g_sM$ precisely these subleading corrections become relevant.
    One should therefore extend the calculations to the curved KS background. This could also enable one to see the KPV decay explicitly from the brane stack perspective.  

    \item We have only taken into account the next-to-leading order commutator corrections in the Myers action. In order to access the phenomenologically interesting regime of larger $p/M$ one should sum up all orders in commutators which is in principle possible.
    
    \item It would be valuable to obtain a clear match in the regime of large $p$ between the on-shell potential of the NS5-brane and $\overline{D3}$-brane picture at higher order in $\alpha'$. A possible way how this can be done is by formulating the potential in both perspectives in terms of $R_{S^2}$.
    In case of the D0-D2-system this has also been done along these lines in \cite{Myers:1999ps} where the potentials can be matched at leading order in commutators and $\alpha'$. This does also work for the $\overline{D3}$-NS5-setup but is difficult when trying to include higher order commutators on the $\overline{D3}$ side. These are already included in the NS5-brane picture. A clear match of the tree level KPV NS5-brane potential should be possible when taking into account the two previous bullet points on the $\overline{D3}$-side, namely the curved KS background and all higher orders in commutators. When it comes to $\alpha'$ corrections, the main obstacles of the matching are twofold. First, not all relevant $\alpha'^2$ terms are known. In particular terms of the form $(\nabla F_5)^2$ on D3-branes appear to be important but are currently not known. Second, the extrinsic curvature corrections present and crucial on the NS5-brane are special. From the brane stack perspective the extrinsic curvature $\Omega$ is only seen indirectly via the derivatives of the nonabelian scalars but does not enter in the calculation of the static points of the potential:
    \begin{equation}
        \Omega^\mu_{\alpha \beta} = \partial_\alpha \partial_\beta Y^\mu - (\Gamma_T)^\gamma_{\alpha \beta} \partial_\gamma Y^\mu + \Gamma^\mu_{\nu \rho} \partial_\alpha Y^\nu \partial_\beta Y^\rho\,,
    \end{equation}
    where $Y^i = 2\pi \alpha' \Phi^i$. We leave it for future investigation to work out a clear matching of both perspectives of the Myers effect beyond the leading order. 
    \item Our analysis has remained at the probe level since we have neglected any backreaction effects of the stack of $\overline{D3}$-branes which is an important subject in the discussion of KPV. For a summary of the status on this subject we refer to the review \cite{VanRiet:2023pnx}.

    \item As it is commonly done, we have set $p=1$ to derive the weakest bounds on $g_sM^2$. However, it is unclear how a single brane should puff up into a fuzzy $S^2$ . The proper interpretation as a fuzzy $S^2$ would require $p\gg 1$. Additionally, the nonabelian features of the brane stack are only present for $p>1$.
\end{itemize}

\section{Conclusions}\label{sec:conclusion}

In this work we have studied a nonabelian $\overline{D3}$-brane stack at the tip of the KS throat at higher order in $\alpha'$. This extends the leading order result of \cite{Kachru:2002gs} by higher order commutators as well as all currently known $\alpha'$ corrections to $\overline{D3}$-branes. This complements the higher order in $\alpha'$ analysis of \cite{Hebecker:2022zme,Schreyer:2022len} of the NS5-brane at the tip of the throat. The reason is that the stack of $\overline{D3}$-branes and NS5-brane picture are dual descriptions of the same system valid in different regimes of parameter space. 
In particular, from the $\overline{D3}$-brane perspective we are able to partially access the regime where the radius of the $S^2$ wrapped by the NS5-brane is string size which has been the main limitation of the NS5-brane analysis \cite{Hebecker:2022zme,Schreyer:2022len}. 

The main result of this work is the $\alpha'^2$ corrected scalar potential \eqref{eq:Vcorr} for the stack of $\overline{D3}$-branes including next-to-leading order commutator corrections as well as higher derivative corrections. From this scalar potential we derive the $\alpha'$ and commutator corrected radius of the fuzzy $S^2$ given by \eqref{eq:rs2corrected}, and analyze in which region of the $(g_sM,p/M)$ parameter space a metastable SUSY breaking minimum, which is necessary for the $\overline{D3}$-brane uplift to work, exists. As a necessary criterion for the minimum to exist, we use that the radius of the fuzzy $S^2$ should be small compared to the radius of the $S^3$ at the tip of the throat, i.e.~$R_{S^2}/R_{S^3}\equiv c_m \ll1$. 
Depending on how small one wishes the ratio $c_m$ to be yields different lower bounds on $g_sM^2$ ranging from 80 to 300 for $p=1$ (see Tab.~\ref{tab:gsm2}) in regions of parameter space where we do also trust the approximations which went into our calculations.
Since our analysis is only a first step beyond the leading order result, these numbers should only be taken as a proxy. Nevertheless, our findings support the important result of \cite{Hebecker:2022zme,Schreyer:2022len}: Higher order corrections have a significant impact on the KPV decay process in the phenomenological relevant regime of parameter space where $g_sM^2$ is as small as possible. A working $\overline{D3}$-brane uplift therefore requires a large contribution to the D3 tadpole from within the throat. 

We have also found that the new way of uplifting proposed in \cite{Hebecker:2022zme,Schreyer:2022len} in which $\alpha'$ corrections of the scalar potential are tuned against the tree level contributions can only work in the regime of small $g_sM$. This means that we can exclude the mechanism for large $g_sM$ and small $p/M$ since the scalar potential of the stack of $\overline{D3}$-brane is strictly positive in this regime. 

As discussed in Sect.~\ref{sec:comparison}, there are three important limitations of the results of this work. First, in the nonabelian calculations we assumed the background in which the $\overline{D3}$-stack expands to be flat and not given by the KS geometry. Second, we have only taken into account the next-to-leading order commutator corrections and neglected all others. Third, we are only able to take into account all currently known $\alpha'$ corrections to D-branes. Extending the results in each of these directions would be important to study and control the $\overline{D3}$-brane stack perspective for larger values of $p/M$ and smaller values of $g_sM$ which is the phenomenologically interesting regime of parameter space. 

Finally, we have not been able to provide a precise matching between the on shell scalar potential from both perspectives. 
The way forward here appears to be a reformulation of both potentials in terms of $R_{S^2}$. Even neglecting $\alpha'$ corrections to the NS5-brane scalar potential, matching the commutator corrections to the leading order in $\alpha'$ KPV NS5-brane potential is challenging. 
The precise matching requires summing up all commutator corrections and to go beyond the flat background approximation. Performing these steps is valuable as it would deepen the understanding of the KPV setup, and also more generally the properties of nonabelian brane stacks and the Myers effect at higher order in $\alpha'$.
We leave this interesting task for future work.

\section*{Acknowledgments}

I thank Arthur Hebecker, Ruben Küspert, Severin Lüst, and Gerben Venken for valuable discussions. I thank Ruben Küspert and Severin Lüst for comments on a draft of this manuscript.

\appendix

\section{$\boldmath{\alpha'}$ corrections to $\boldmath{\overline{D3}}$-branes at the tip of the KS throat} \label{sec:app}

In this Appendix, we elaborate on the details of $\alpha'$ corrections to $\boldmath{\overline{D3}}$-branes at the tip of the KS throat. We take into account the same set of $\alpha'$ corrections as in \cite{Hebecker:2022zme,Schreyer:2022len}, namely all corrections derived in \cite{Bachas:1999um,Garousi:2009dj,Garousi:2010ki,Garousi:2010rn,Garousi:2011ut,Robbins:2014ara,Garousi:2014oya,Jalali:2015xca,Garousi:2015mdg,Jalali:2016xtv,BabaeiVelni:2016srs,Garousi:2022rcv}.

Before evaluating all non-vanishing $\alpha'$ corrections for our setup, we note that the curvature corrections to the CS action of a stack of $\overline{D3}$-branes
\begin{equation}
    S_{\text{CS,}\overline{D3}}=T_3 \int \text{STr} \left( P \left[ \text{e}^{i\lambda \iphi\iphi} \left( \sum B^{(n)} \wedge \text{e}^{-C_2} \right)\right]\wedge\text{e}^{\lambda F} \wedge \sqrt{\frac{\hat{A}(2\pi \lambda R_T)}{\hat{A}(2\pi \lambda R_N)}}\right)\,,
    \label{eq:csdptreelevel}
\end{equation}
do not contribute in our setup. The reason is that when expanding the A-roof genus term in \eqref{eq:csdptreelevel} up to $\mathcal{O}(\alpha'^2)$, the curvature term is a 4-form along the $\overline{D3}$-brane. The index structure is then such that all terms including the curvature term will vanish because the nonabelian interior product will either act on $C_4$ or the curvature term but these necessarily only have indies along the brane. The nonabelian interior product with these terms hence vanishes. 

The in our case non-vanishing terms contributing to the DBI action are given by~\cite{Bachas:1999um,Robbins:2014ara,Garousi:2014oya}
\begin{equation}
    S_\text{DBI} = -T_p \int \dd^{4}\sigma \,\text{STr} \left(\text{e}^{-\phi}\sqrt{\det(Q^i_{~j})} \sqrt{-\det\left( T_{ab} \right)}\left(1+
    \frac{\pi^2\alpha'^2}{48}\mathcal{L}_{\alpha'^2}\right)\right)
\end{equation}
where $T_{ab}= P[E_{ab} + E_{ai}(Q^{-1}-\delta)^{ij} E_{jb}] +\lambda F_{ab} $ and 
\begin{equation}
\label{eq:dbinonvanishing}
    \mathcal{L}_{\alpha'^2} = - 2 \hat{R}_{ij}\hat{R}^{ij} +\frac{1}{24} H^{ijk}H_i^{~lm} H_{jl}^{~~n} H_{kmn} + \frac{g_s^4}{24} F^{ijk}F_i^{~lm} F_{jl}^{~~n} F_{kmn}+ \frac{1}{2} H^{ijk} H_{ij}^{~~l} \hat{R}_{kl} \,,
\end{equation}
where\footnote{Note that the second fundamental Form does not appear in the pulled-back Riemann tensor since the branes are pointlike in the internal space.} $\hat{R}_{ij}=R^a_{~iaj}$, and the terms including $F_3$ have been inferred from the $H_3^4$ terms by S-duality of the (anti)-D3-brane action and, to the best of the author's knowledge, do not appear anywhere in the literature. 
The correct $g_s$ dependence of the $F_3^4$ term can be obtained by making use of the non-holomorphic Eisenstein series $E_1(\tau,\bar{\tau})$ following the argument given in App.~C.2 of \cite{Schreyer:2022len}. 

Next, we evaluate these terms for our setup. For this, we use the parametrization of the fluxes given by \cite{Klebanov:2000hb}
\begin{align}
    \label{eq:h3}
    H_3 \supset &\,\frac{g_s M \alpha' }{6}  \dd\tau\wedge g^3\wedge g^4 + \frac{g_s M \alpha' \tau}{12} g^5\wedge\left(g^1\wedge g^3 + g^2\wedge g^4\right)+\mathcal{O}(\tau^2)\,,\\
    \label{eq:f3}
    F_3  \supset& \,\frac{M\alpha'}{2} g^5\wedge g^3 \wedge g^4+ \frac{M \alpha' \tau}{12}\dd\tau\wedge\left(g^1\wedge g^3 + g^2\wedge g^4\right)+ \mathcal{O}(\tau^2)\,,
\end{align}
where the 2-forms $g^{1,\dots, 5}$ parametrize the $T^{1,1}$ and are for instance given in \cite{Klebanov:2000hb}. 
Since we are not interested in the dynamics of the system but only in the extrema of the potential we can set $T_{ab} = g_{ab}$ and then evaluate \eqref{eq:dbinonvanishing} at the tip of the KS throat using the KS metric. One finds 
\begin{equation}
\label{eq:dbinonvanishingevaluated}
\begin{split}
   \frac{\pi^2\alpha'^2}{48} \mathcal{L}_{\alpha'^2} = &\frac{\pi^2\alpha'^2}{48 (g_sM)^2 I(0)^3}\left(-24\times 6^{2/3} I''(0)^2+\frac{1}{36}+\frac{9}{4}+ \frac{2\times2^{1/3}I''(0)}{3^{2/3}}\right)\\
   =& -\frac{4.9059}{(g_sM)^2} \equiv -\frac{c_1}{(g_sM)^2}\,,
   \end{split}
\end{equation}
where we used that $I(0)\approx 0.71805$ and $I''(0)=-2^{2/3}/3^{4/3}$ are the warp factor integral and its second derivative evaluated at the tip of the throat. 

The only non-vanishing correction to the CS action for our setup is \cite{Garousi:2010ki}
\begin{equation}
\label{eq:csnonvanishing}
    S_{\text{CS},\alpha'^2} \supset  \frac{\pi^2\alpha'^2T_3}{24}\frac{2}{(p+1)!}\int \dd^4x\,\Str\left(\epsilon^{a_0a_1a_2a_3} D_j \tilde{F}^{(5)}_{i a_0a_1a_2a_3} \hat{R}^{ij}\right)\,,
\end{equation}
where $\epsilon^{a_0a_1a_2a_3}$ is the Levi-Civita symbol on the worldvolume of the brane. To evaluate \eqref{eq:csnonvanishing} in our setup we use 
\begin{equation}
    \label{eq:f5}
    \tilde{F}_5  \supset \left(\frac{\tau}{3^{4/3} g_s^3 M^2 I(0)^2}+\mathcal{O}(\tau^3)\right) \dd x^0 \wedge \dd x^1 \wedge \dd x^2 \wedge \dd x^3 \wedge \dd\tau+\mathcal{O}(\tau^2)\,,
\end{equation}
and obtain
\begin{equation}
    \label{eq:csnonvanishingevaluated}
    S_{\text{CS},\alpha'^2} \supset - \frac{T_3 p}{g_s} \int \dd^4x \sqrt{-g_4} \frac{c_2}{(g_sM)^2}\,,
\end{equation}
with $c_2\approx 31.5953$.


\bibliographystyle{JHEP}
\bibliography{refs}

\end{document}